\documentclass[11pt, a4paper]{article}
\usepackage{amsmath}
\usepackage{amsfonts}
\usepackage{amssymb}
\usepackage{epsfig}
\usepackage{amsthm}
\usepackage{multirow}
\usepackage{color}
\usepackage{algorithm}
\usepackage{algorithmic}

\begin{document}

\title{\LARGE \LARGE \LARGE  Constructing Intrinsic Delaunay Triangulations from the Dual of Geodesic Voronoi Diagrams \\[0.5ex]}
\author{Yong-Jin Liu$^1$, Chun-Xu Xu$^1$, Dian Fan$^1$, Ying He$^2$}
\date{\normalsize $^1$ Tsinghua National Lab for Information Science and Technology,\\
Department of Computer Science and Technology, \\
Tsinghua University, Beijing, P. R. China\\
$^2$ School of Computer Engineering\\
Nanyang Technological University, Singapore
}

\maketitle

\begin{abstract}

\normalsize \noindent \setlength{\baselineskip}{15pt}Intrinsic Delaunay triangulation (IDT) is a fundamental data structure in computational geometry and computer graphics.
  However, except for some theoretical results, such as existence and uniqueness, little progress has been made towards computing IDT on simplicial surfaces.
  To date the only way for constructing IDTs is the edge-flipping algorithm,
  which iteratively flips the non-Delaunay edge to be locally Delaunay.
  Although the algorithm is conceptually simple and guarantees to stop in finite steps, it has no known time complexity.
  Moreover, the edge-flipping algorithm may produce non-regular triangulations, which contain self-loops and/or faces with only two edges.
  In this paper, we propose a new method for constructing IDT on manifold triangle meshes.
  Based on the duality of geodesic Voronoi diagrams, our method can guarantee the resultant IDTs are regular.
  Our method has a theoretical worst-case time complexity $O(n^2\log n)$ for a mesh with $n$ vertices.
  We observe that most real-world models are far from their Delaunay triangulations,
  thus, the edge-flipping algorithm takes many iterations to fix the non-Delaunay edges.
  In contrast, our method is non-iterative and insensitive to the number of non-Delaunay edges.
  Empirically, it runs in linear time $O(n)$ on real-world models.

  As a by-product, the regular Delaunay triangulations naturally induce discrete Laplace-Beltrami operators (LBOs),
  which are intrinsic to the geometry and have non-negative weights.
  We evaluate the commonly used discrete LBOs on the original triangulations and the intrinsic Delaunay triangulations.
  Computational results show that the IDT induced LBOs are more accurate than the LBOs defined on the original mesh.
  Moreover, their discrete Laplacian matrices have smaller condition number than the original triangulations.
  As a result, IDTs are ideal for applications which solve the linear system and eigensystem of the discrete Laplacian.
\\[1ex]



\end{abstract}

\pagebreak \setlength{\baselineskip}{20pt}
\setlength{\topmargin}{-1cm}

\section{Introduction}
\label{sec:intro}
  A Delaunay triangulation for a set $\bf P$ of points in $\mathbb{R}^2$ is a triangulation such that no point in $\bf P$ is inside the circumcircle of any triangle in the triangulation.
  It is well known that Delaunay triangulations tend to avoid skinny triangles,
  since they maximize the minimum angle of all the angles of the triangles in the triangulation.
  Although the Delaunay triangulations in Euclidean spaces are well understood~\cite{Okabe2000},
  the \textit{intrinsic} Delaunay triangulations on Riemannian manifolds have received less attention.
  By using the closed ball property~\cite{Edelsbrunner1997},
  Dyer et al.~\cite{Dyer2008} proposed adaptive sampling criteria for constructing intrinsic Voronoi diagram and its dual Delaunay triangulation on 2-manifolds.
  Recently, Boissonnat et al.~\cite{Boissonnat2013} proposed an algorithm for constructing intrinsic Delaunay triangulation on smooth closed submanifold of Euclidean space.
  Both methods are based on the \textit{convex neighborhood}, which, in general, is an extremely small region around a point on the surface.
  As a result, in spite of their important theoretical values, they are not practical for piecewise linear surfaces,
  which are dominant in digital geometry processing.

  Rivin~\cite{rivin1994} and Indermitte et al.~\cite{Indermitte2001} defined intrinsic Delaunay triangulation (IDT) on triangle meshes,
  where the IDT edges are geodesic paths and the \textit{geodesic} circumcircles of all Delaunay triangles have empty interiors.
  The intrinsic Delaunay triangulation has many nice properties.
  For example, Bobenko and Springborn~\cite{Bobenko2007} proved that the classic cotangent Laplace-Beltrami operator (LBO) has non-negative weights $w_{ij}$,
  if and only if the underlying triangulation is Delaunay.
  They also proposed a new LBO which depends only on the intrinsic geometry of the surface and its edge weights are non-negative.

  Using intrinsic properties, such as the strong convexity radius and the injectivity radius,
  Dyer et al.~\cite{Dyer2008} presented adaptive sampling criteria which guarantee the IDT is regular.
  Their elegant results establish inequalities that relate these intrinsic properties to the local feature size.
  However, to our knowledge, there is no practical algorithm for computing strongly convex regions on meshes.
  To date, the only practical algorithm for computing IDT is the edge-flipping algorithm~\cite{Bobenko2007,Fisher2006,Indermitte2001},
  which iteratively flips the non-locally Delaunay edge to be locally Delaunay.
  The edge-flipping algorithm is conceptually simple and easy to implement.
  Indermitte et al.~\cite{Indermitte2001} showed that the edge flipping algorithm terminates in a finite number of steps and thus the intrinsic Delaunay tessellation exists.
  Bobenko and Springborn~\cite{Bobenko2007} proved the uniqueness of the intrinsic Delaunay tessellation.
  Although the edge flipping algorithm converges, it has no known time complexity.
  Moreover, the resultant IDT may contain \textit{non-regular} triangles,
  which have either self-loops (i.e., edges with identical end points) or faces with only two edges.

  In this paper, we propose a new method for constructing intrinsic Delaunay triangulation on 2-manifold triangle meshes.
  Our idea is to compute a geodesic Voronoi diagram (GVD), which has a regular dual triangulation.
  Given a 2-manifold mesh with $n$ vertices, our method first computes the GVD by taking all vertices as sites.
  Then, our method identifies the Voronoi cells that violate the closed ball property~\cite{Edelsbrunner1997} and fixes them by adding auxiliary sites.
  We show that by adding at most $O(n)$ sites, the dual graph of the geodesic Voronoi diagram is an intrinsic Delaunay triangulation,
  which is guaranteed to be \textit{regular}.
  Moreover, thanks to the bounded time complexity of computing geodesic Voronoi diagrams~\cite{Liu2011},
  our method has a theoretical worst-case time complexity $O(n^2\log n)$ .
  We observe that most real-world models are far from their Delaunay triangulations,
  thus, the edge-flipping algorithm takes many iterations to converge,
  whereas our method is not sensitive to the number of non-Delaunay edges and it \textit{empirically} runs in linear time $O(n)$ on these models.

  As a by-product, the regular Delaunay triangulations naturally induce discrete Laplace-Beltrami operators (LBOs),
  which are intrinsic to the geometry and have non-negative weights.
  We evaluate the commonly used discrete LBOs on the original triangulations and the intrinsic Delaunay triangulations.
  Computational results show that the IDT induced LBOs are more accurate than the LBOs defined on the original mesh.
  Moreover, their discrete Laplacian matrices have smaller condition number than the original triangulations.
  As a result, IDTs are ideal for applications which solve the linear system and eigensystem of the discrete Laplacian.
  We demonstrate the IDTs on harmonic mapping and manifold harmonics,
  and observe that the IDTs produce more accurate and robust results than the original triangulations.

  The remaining of the paper is organized as follows:
  Section~\ref{sec:preliminaries} reviews the mathematical background and highlights the fundamental difference between the 2D Delaunay triangulation and the IDT on meshes.
  Section~\ref{sec:IDT} details our algorithm for computing IDTs,
  followed by the experimental results and comparison in Section~\ref{sec:results}.
  Section~\ref{sec:LBO} shows the IDT induced discrete LBOs are more accurate and robust than the original meshes,
  and demonstrates them on solving the linear system and eigensystem of the discrete Laplacian.
  Finally, Section~\ref{sec:conclusion} concludes the paper.
  To ease reading, we list the main notations in Table~\ref{table:notation} and delay the long proofs in Appendix.

  \begin{table}
  \begin{center}
  \begin{tabular}{|l|l|}
  \hline
  $M$ & a manifold triangle mesh\\
  $V,E,F$  & the vertex, edge and face sets of $M$\\
  $v,v_i,v_j,\cdots$ & vertex\\
  $e,e_{ij}=\{v_i,v_j\},\cdots$ & edge\\
  $f,f_{ijk}=\{v_i,v_j,v_k\},\cdots$ & triangular face\\
  $p,q, p_i,q_i, \cdots$ & point on $M$\\
  $\gamma(p,q)$ & the geodesic path between $p$ and $q$\\
  $d(p,q)$ & the geodesic distance between $p$ and $q$\\
  $n (=|V|)$ & number of vertices\\
  \hline
  \hline
  $GVD(P)$ & the geodesic Voronoi diagram of sites $P$\\
  $\Upsilon$ & the set of Voronoi vertices\\
  $B$ & the set of Voronoi edges\\
  $C$ & the set of Voronoi cells\\
  $VC(p_i)$ & the Voronoi cell of site $p_i$\\
  $b(p,q)$ & the bisector of sites $p$ and $q$\\
  $pb(p)$ & the pseudo bisector of site $p$ in Section~\ref{subsec:homemorphism}\\
  $V' (\supseteq V)$ & the set of augmented sites in Section~\ref{subsec:homemorphism}\\
  $V'' (\supseteq V')$ & the set of augmented sites in Section~\ref{subsec:intersection}\\
  $V''' (\supseteq V'')$ & the set of augmented sites in Section~\ref{subsec:boundary}\\
  \hline
  \hline
  $IDT(M)$ & the intrinsic Delaunay triangulation of $M$\\
  $\Xi$   & the set of \textit{g}-edges\\
  $\Gamma$ & the set of geodesic triangles\\
  $\xi$ & \textit{g}-edge\\
  $\tau$ & geodesic triangle\\
  \hline
  \end{tabular}
  \end{center}
  \caption{Main notations.}
  \label{table:notation}
  \end{table}

  \begin{figure}[th]
  \centering
  \includegraphics[width=0.13\textwidth, bb=131 352 434 670, clip=true]{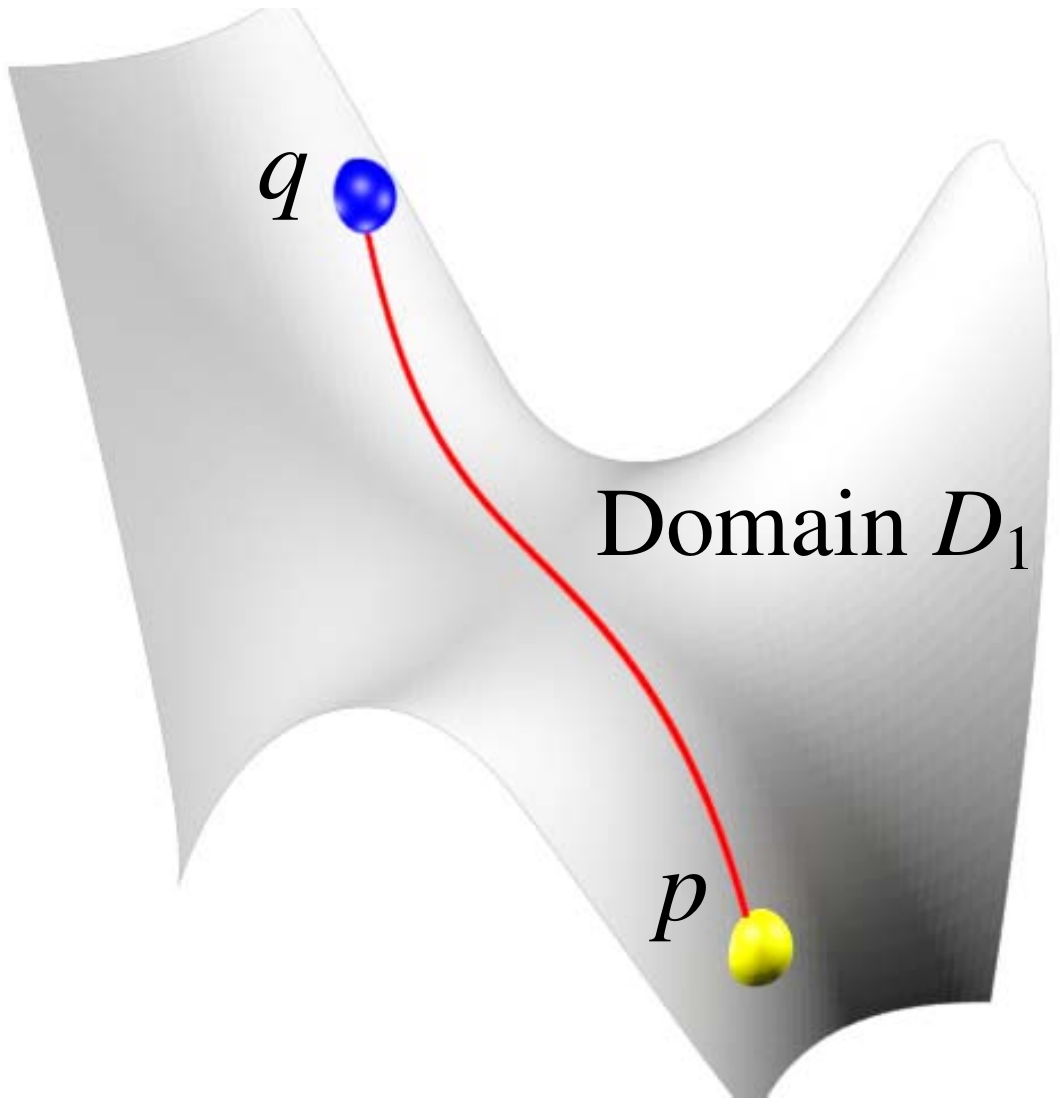}\hspace{12pt}
  \includegraphics[width=0.19\textwidth, bb=142 426 421 622, clip=true]{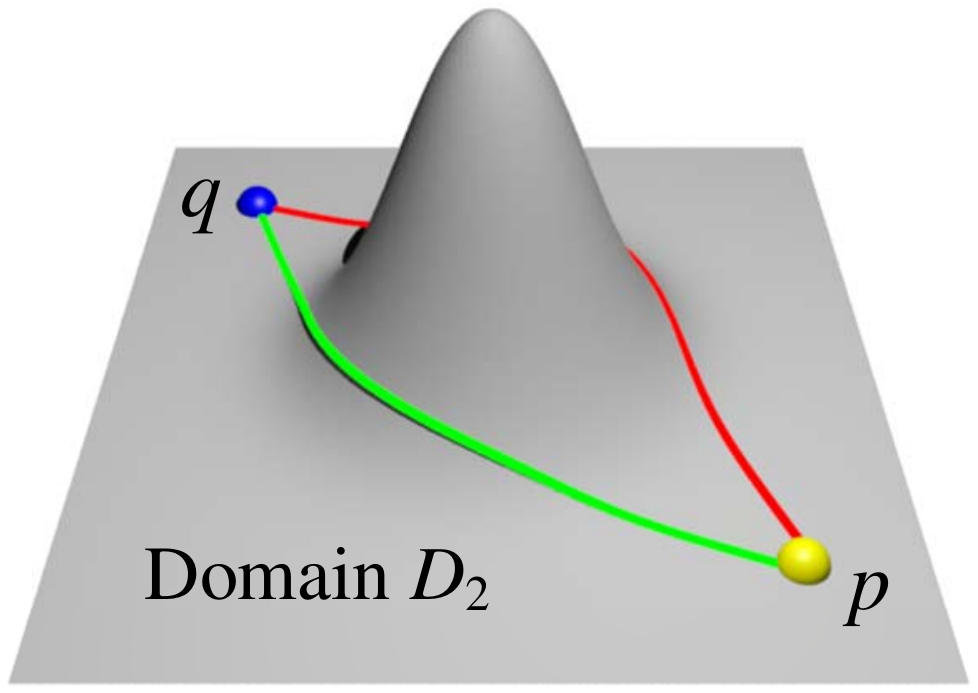}\hspace{12pt}
  \includegraphics[width=0.19\textwidth, bb=141 448 423 579, clip=true]{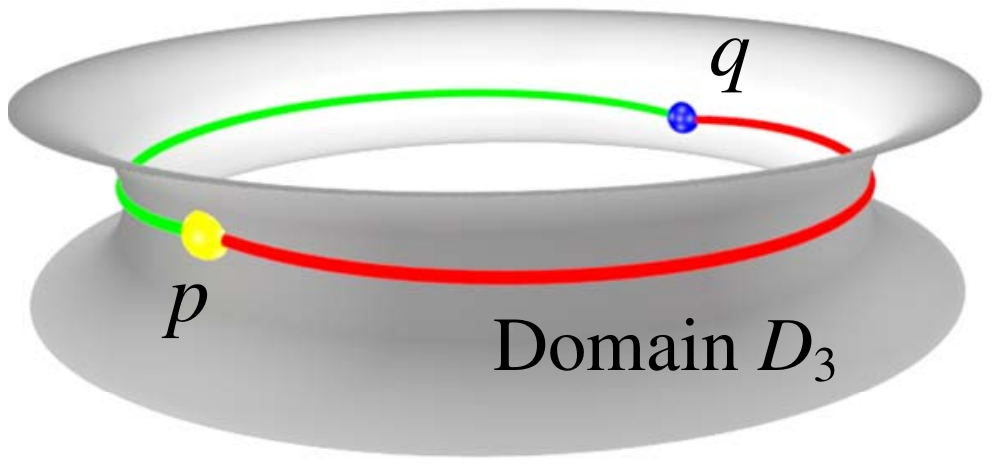}\hspace{12pt}
  \includegraphics[width=0.19\textwidth, bb=142 406 424 633, clip=true]{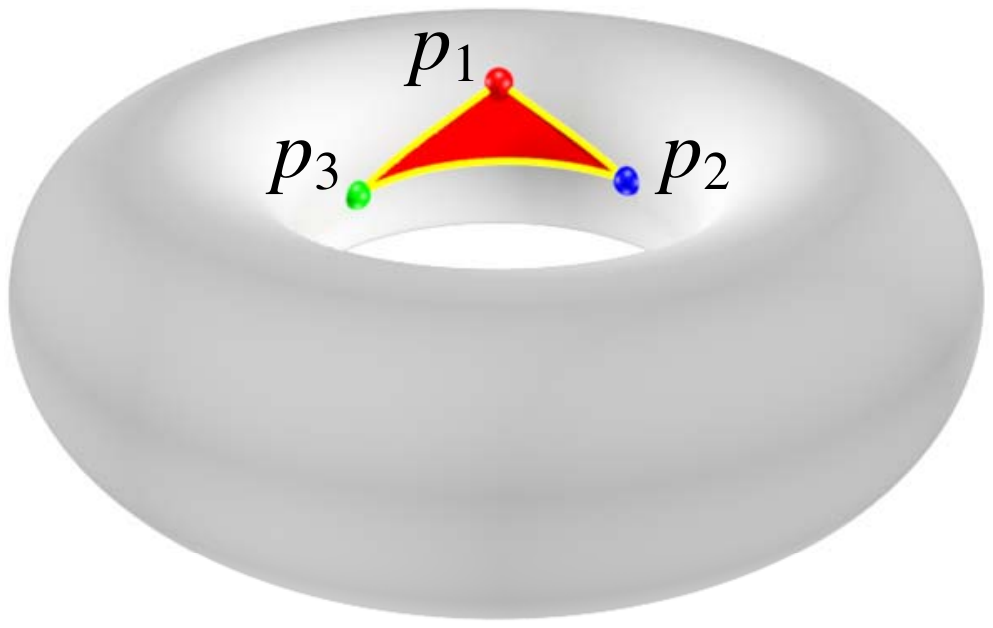}\hspace{12pt}
  \includegraphics[width=0.13\textwidth, bb=169 361 398 593, clip=true]{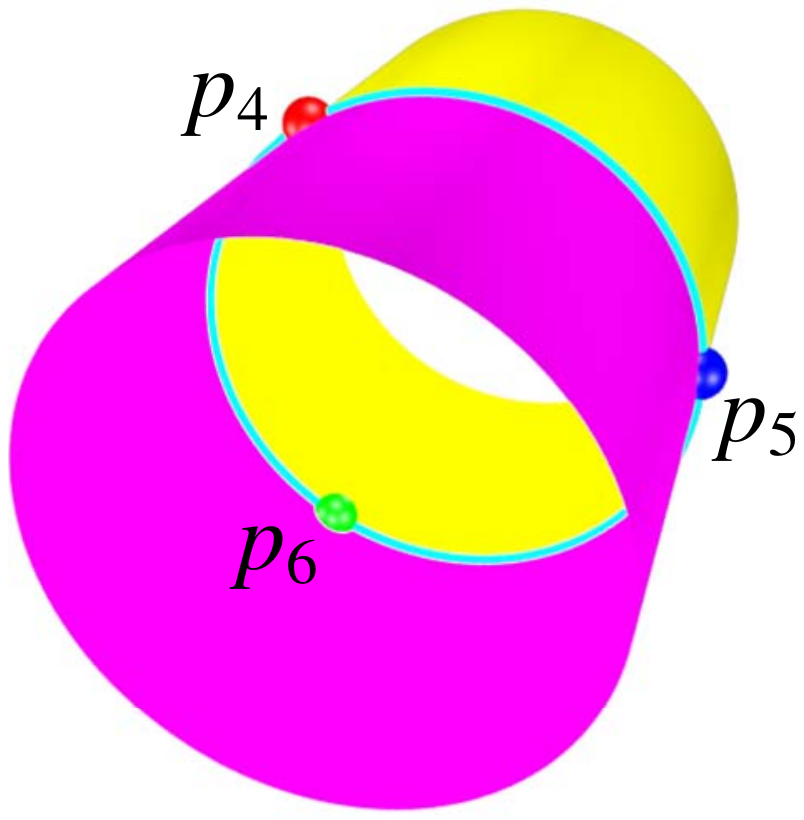}\\
  \makebox[0.7in]{(a)}\makebox[1.4in]{(b)}\makebox[1.4in]{(c)}\makebox[1.2in]{(d)} \makebox[1.1in]{(e)}\\
  \vspace{-0.1in}
  \caption{Geodesic paths \& geodesic triangles. (a) $D_1$ is a simply connected domain with non-positive Gaussian curvature everywhere.
  There is a \textit{unique} geodesic path between any pair of distinct points in $D_1$. (b) $D_2$ is also simply connected.
  However, it has a part with positive Gaussian curvature. As a result, there are two geodesic paths between $p$ and $q$.
  (c) $D_3$ has negative Gaussian curvature everywhere. However, as its fundamental group is non-trivial, the geodesic paths are not unique.
  (d) The red region $\tau=(p_1,p_2,p_3)$ is a geodesic triangle.
  However, its complement $M\setminus\tau$, although having three geodesic sides, is not a geodesic triangle, since it is of genus 1.
  (e) $\gamma(p_4,p_5)$, $\gamma(p_5,p_6)$ and $\gamma(p_6,p_4)$ are three geodesic paths wrapping the cylinder.
  Each colored region has three geodesic sides. However, none of them is a geodesic triangle, since it is multiply connected.
  }
  \label{fig:geodesic_path}
  \end{figure}

\section{Preliminaries}
\label{sec:preliminaries}

  Let $M$ be a manifold triangle mesh, and $V$, $E$, $F$ be the set of vertices, edges and faces of $M$.
  Every interior point of $M$ has a neighborhood which is isometric to either a neighborhood of the Euclidean plane or a neighborhood of the apex of a Euclidean cone.

\subsection{Geodesic Paths, Geodesic Triangles and Geodesic Circumcircles}
  Consider two points $p,q\in M$. A geodesic path between $p$ and $q$, denoted by $\gamma(p,q)$, is the \textit{locally} shortest path between them.
  Mitchell et al.~\cite{Mitchell1987} showed that the general form of a geodesic path is an alternating sequence of vertices and (possibly empty) edge sequences
  such that the unfolded image of the path along any edge sequence is a straight line segment
  and the angle of $\gamma$ passing through a vertex is greater than or equal to $\pi$.
  The vertices with cone angles more than $2\pi$ are called \textit{saddle vertices}, which play a critical role in geodesic computation~\cite{Ying13}.
  We denote by $d(p,q)$ the geodesic distance between $p$ and $q$, and $b(p,q)$ the bisector of $p$ and $q$.

  Given a simply connected domain $\Omega$ with negative Gaussian curvature everywhere, the geodesic path $\gamma(p,q)$ is unique for any pair of points $p,q\in\Omega$.
  However, in general, geodesics are not unique for regions with positive Gaussian curvature and/or non-trivial topology.
  See Figure~\ref{fig:geodesic_path}(a)-(c).

  \begin{figure}[th]
  \centering
  \includegraphics[width=0.99\textwidth, bb=29 549 564 679, clip=true]{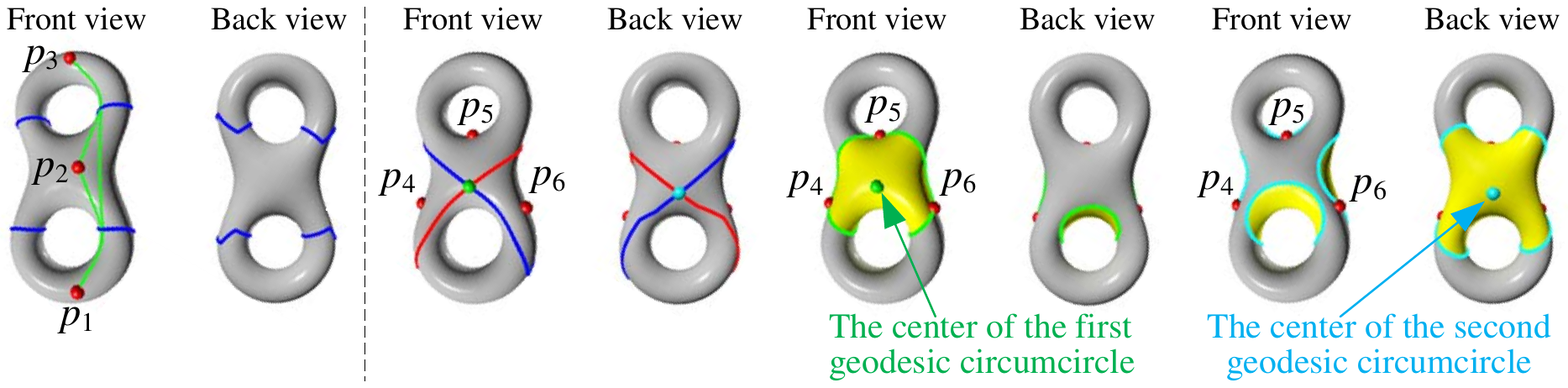}\\
  \makebox[0.8in]{(a) $\bigtriangleup p_1p_2p_3$ has no circumcircle}
  \makebox[4.1in]{(b) $\bigtriangleup p_4p_5p_6$ has 2 circumcircles}
  \vspace{-0.1in}
  \caption{Geodesic circumcircles. Geodesic triangle $\bigtriangleup p_1p_2p_3$ has no geodesic circumcircle,
  since the bisector $b(p_1,p_2)$ (shown in blue) does not meet $b(p_2,p_3)$.
  In contrast, geodesic triangle $\bigtriangleup p_4p_5p_6$ has two geodesic circumcircles, since $b(p_4,p_5)$ (blue) and $b(p_5,p_6)$ (red) intersect twice.}
  \label{fig:geodesic_circumcircle}
  \end{figure}

  \ \\
  \noindent\textbf{Definition 1} (Geodesic Triangle). \textit{A geodesic triangle $\tau\in M$ is a simply connected domain whose boundary $\partial\tau$ has three geodesic paths.
  Each geodesic path is called a \textit{g}-edge and the endpoints of a \textit{g}-edge are called \textit{g}-vertices.}
  \ \\

  On $\mathbb{R}^2$, any three non-parallel lines form a triangle. However, not all three geodesics on a mesh form a triangle.
  See Figure~\ref{fig:geodesic_path}(d)-(e).

  \ \\
  \noindent\textbf{Definition 2} (Geodesic Disk \& Geodesic Circumcircle). \textit{A geodesic disk centered at a point $p\in M$ of radius $r$,
  denoted by $D(p,r)$, consists of all points whose distance to $p$ does not exceed $r$, i.e., $D(p,r)=\{q\in M:d(p,q)\leq r\}$.
  If a geodesic disk is simply connected, its boundary $\partial D(p,r)$ is called geodesic circumcircle.}
  \ \\

  On $\mathbb{R}^2$, each non-degenerate triangle has a unique circumcircle that passes through its three vertices.
  However, a geodesic triangle may have no geodesic circumcircle at all or more than 1 geodesic circumcircles.
  See Figure~\ref{fig:geodesic_circumcircle}.

\subsection{Intrinsic Delaunay Triangulations}

  In contrast to Euclidean spaces, Delaunay triangulations do not exist for an arbitrary set of points on a Riemannian manifold.
  Leibon and Letscher~\cite{Leibon2000} proposed sampling density conditions to ensure that the triangulation can accurately represent both the topology and geometry of the manifold.
  However, Boissonnat et al.~\cite{Boissonnat2013} pointed out that sampling density alone is insufficient to guarantee an intrinsic Delaunay triangulation for Riemannian manifolds of dimension 3 and higher.

  The definition of intrinsic Delaunay triangulation on 2-manifold meshes is due to Rivin~\cite{rivin1994},
  who generalized the 2D Delaunay condition (i.e., a circle circumscribing any Delaunay triangle does not contain any other input points in its interior)
  and required the empty \textit{geodesic} circumcircle property.

  \ \\
  \noindent\textbf{Definition 3} (Intrinsic Delaunay Triangulation). \textit{The intrinsic Delaunay triangulation (IDT) on $M$, denoted by $IDT(M)=(V,\Xi,\Gamma)$, is a triangulation such that
  \begin{itemize}
  \item~ the vertex set of $IDT(M)$ equals $V$;
  \item~ every edge $\xi$ in $\Xi$ is a geodesic path on $M$, i.e., a \textit{g}-edge;
  \item~ each face $\tau\in\Gamma$ is a geodesic triangle, which has a geodesic circumcircle containing no mesh vertices in its interior;
  \item~ and $\Gamma$ forms a tessellation of $M$;
  \end{itemize}
  }
  \

  Indermitte et al.~\cite{Indermitte2001} proved the existence by showing the edge-flipping algorithm terminates in finite steps.
  Bobenko and Springborn~\cite{Bobenko2007} proved the uniqueness of Delaunay tessellation (whose faces are generally but not always triangular).
  The Delaunay triangulation can be obtained by triangulating the non-triangular faces.
  They pointed out that the Delaunay triangulation, while in general not unique,
  differs from another Delaunay triangulation only by edges with vanishing cot-weights (i.e., the sum of two angles opposite that edge is $\pi$).

\subsection{Geodesic Voronoi Diagrams}

  \noindent\textbf{Definition 4} (Geodesic Voronoi Diagram). \textit{Let $P=\{p_1,p_2,\dots,p_m\}$ be a set of points on $M$.
  The Voronoi cell $VC(p_i)$ corresponding to site $p_i$ consists of all points whose distance to $p_i$ is less than or equal to their distance to any other site,
  i.e., $VC(p_i)=\{q\in M| d(p_i,q)\leq d(p_j,q),\forall i\neq j\}$.
  The geodesic Voronoi diagram (GVD) of $P$ is the union of all Voronoi cells, $GVD(P)=\{VC(p_1),VC(p_2),\cdots,VC(p_m)\}$.
  The Voronoi edges bounding the Voronoi cells are trimmed bisectors and the Voronoi vertices are points incident to three or more Voronoi edges.}
  \ \\

  \begin{figure}[th]
  \centering
  \includegraphics[width=0.4\textwidth, bb=84 453 496 677, clip=true]{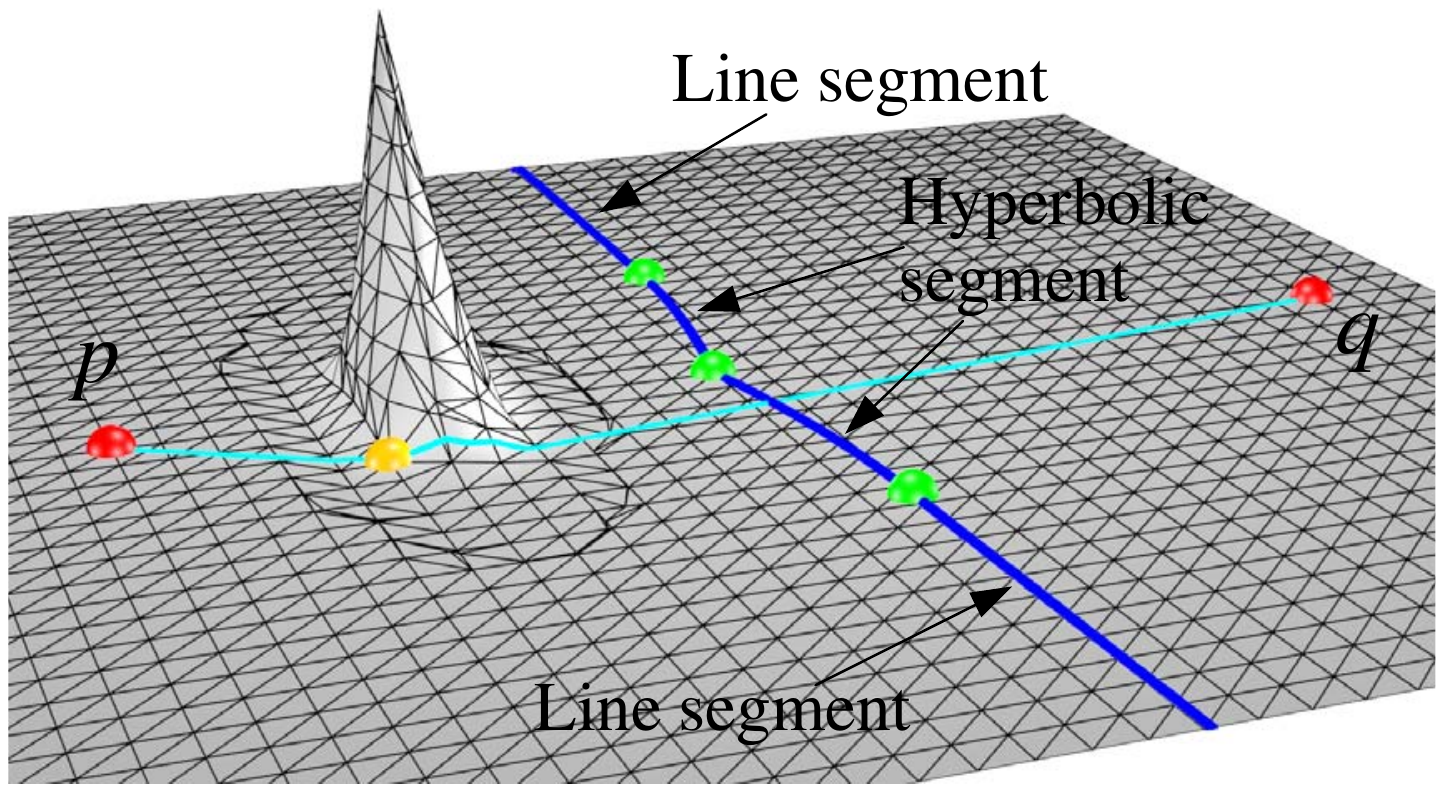}\hspace{20pt}
  \includegraphics[width=0.52\textwidth, bb=16 395 544 591, clip=true]{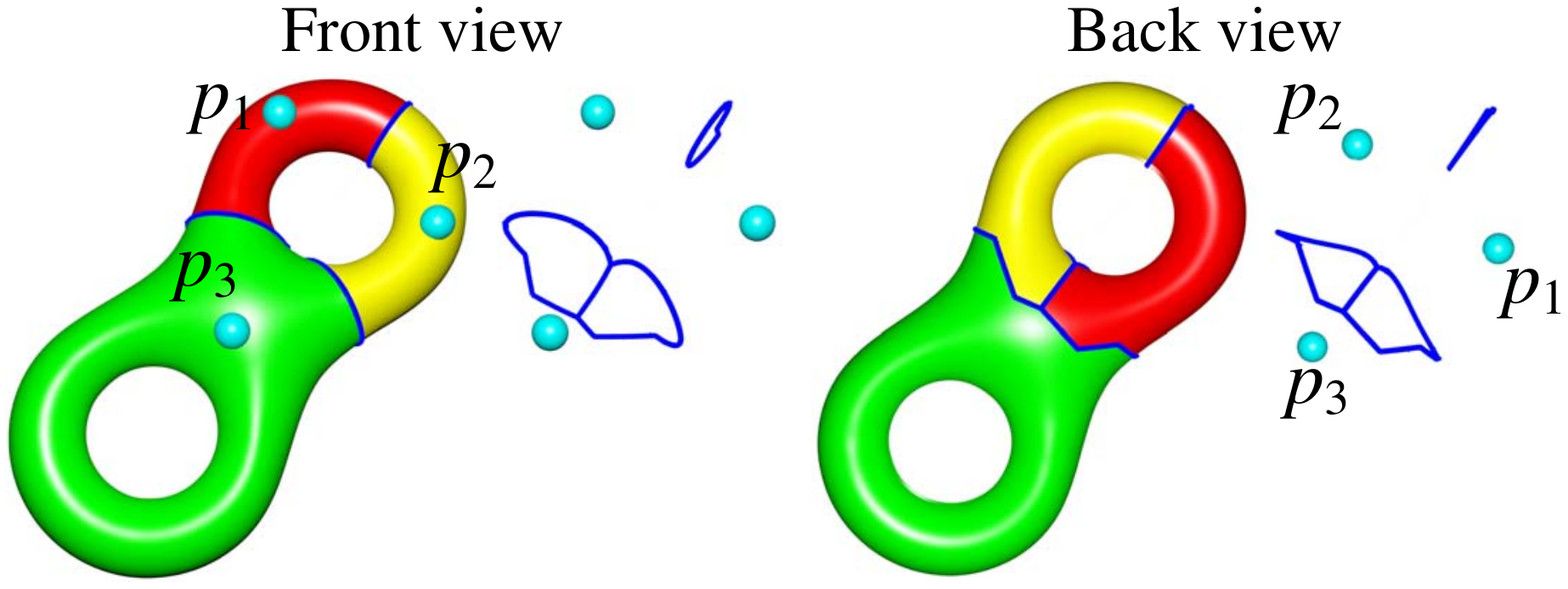}
  \vspace{-0.1in}
  \caption{Bisectors and Voronoi cells on a 2-manifold mesh are significantly different from their Euclidean counterpart.
  Left: There is a saddle vertex (yellow) on the geodesic path $\gamma(p,q)$ (cyan).
  As a result, the bisector $b(p,q)$ (in blue) consists of both line segments and hyperbolic segments.
  Right: $VC(p_3)$ (in green) is of genus-1, and $VC(p_1)$ (in red) and $VC(p_2)$ (in yellow) are of genus-0 but with multiple boundaries.}
  \label{fig:general-VD}
  \end{figure}

  It is shown in~\cite{Liu2011,Xu2014} that the GVD forms a tessellation of $M$,
  since all Voronoi cells are mutually exclusive, and $\bigcup_{i=1}^mVC(p_i)=M$.

  It is well known that Voronoi cells in $\mathbb{R}^2$ are simply connected and convex, and the Voronoi edges are all line segments.
  However, these properties do not hold for geodesic Voronoi diagrams.
  Although a geodesic Voronoi cell is still connected, it may have multiple boundaries and even handles. See Figure~\ref{fig:general-VD} (right).
  Moreover, a geodesic Voronoi cell can be concave, since the bisectors on triangle meshes consist of both line segments and hyperbolic segments~\cite{Liu2011}.
  See Figure~\ref{fig:general-VD} (left).

  \begin{figure}[th]
  \centering
  \includegraphics[width=0.72\textwidth, bb=40 500 517 670, clip=true]{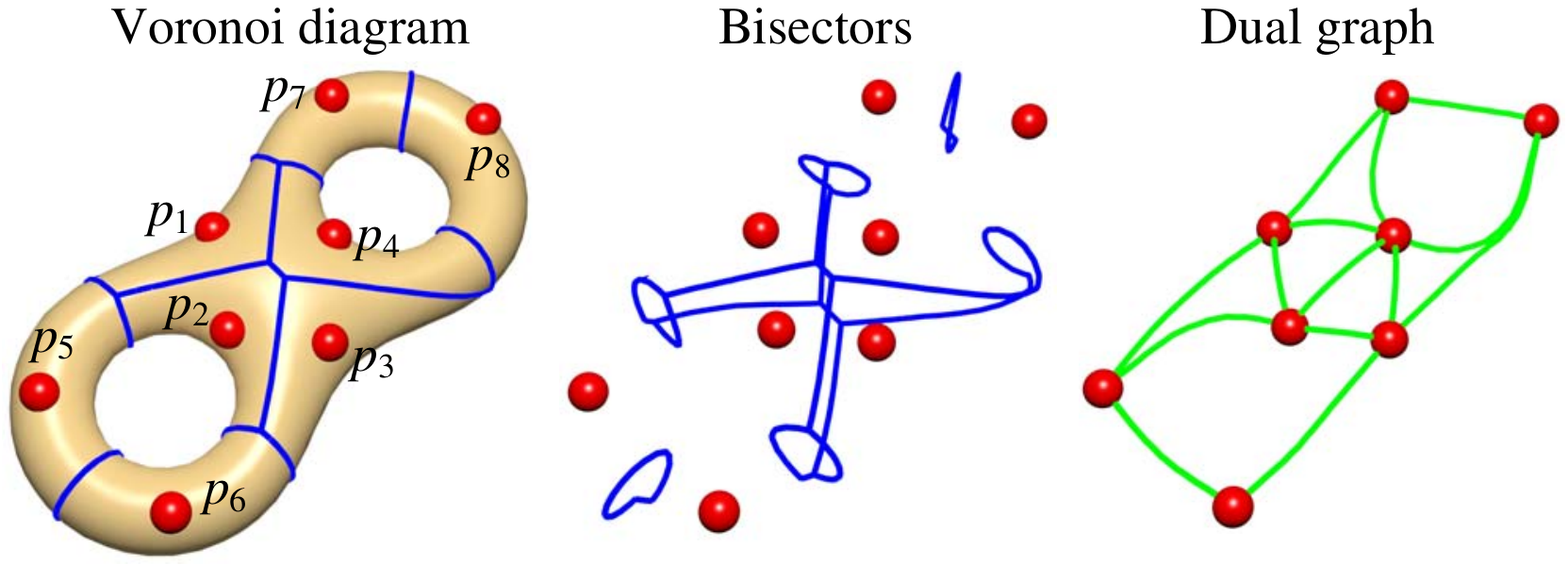}
  \vspace{-0.1in}
  \caption{Consider the Two-hole Torus model with 8 sites $p_i$, $i=1,\cdots,8$.
  Note that $VC(p_5)$, $VC(p_6)$, $VC(p_7)$ and $VC(p_8)$ are multiply connected.
  $VC(p_1)$ and $VC(p_2)$ share 2 edges, and so do the other pairs of Voronoi cells $(VC(p_2),VC(p_3))$, $(VC(p_3),VC(p_4))$, $(VC(p_1),VC(p_4))$ and $(VC(p_2),VC(p_4))$.
  We can clearly see that the dual graph is not a triangulation. In fact, it does not even form a tessellation.}
  \label{fig:VD-dual}
  \end{figure}

\subsection{The Closed Ball Property \& Duality}
\label{subsec:closedball}
  It is well known that the dual graph of a Voronoi diagram in $\mathbb{R}^2$ is the Delaunay tessellation for the same set of points.
  Therefore, one can adopt the Voronoi diagram's algorithm to construct the Delaunay tessellation, and vice versa.
  This duality, however, does not exist on 2-manifold meshes in general.
  See Figure \ref{fig:VD-dual}.

  Edelsbrunner and Shah~\cite{Edelsbrunner1997} introduced the \textit{closed ball property} for triangulating an abstract topological space.
  For a surface without boundary, the closed ball property expresses three conditions:
  \begin{enumerate}
  \item \textbf{Homeomorphism condition}: each Voronoi cell is homeomorphic to a planar disk;
  \item \textbf{2-cell intersection condition}: the intersection of any two Voronoi cells is either empty, or a single Voronoi edge, or a single Voronoi vertex;
  \item \textbf{3-cell intersection condition}: the intersection of any three Voronoi cells is either empty or a single Voronoi vertex.
  \end{enumerate}

  Dyer et al.~\cite{Dyer2008} gave the sufficient condition that a GVD has a dual triangulation.

  {
  \renewcommand{\algorithmicrequire}{\textbf{Input:}}
  \renewcommand{\algorithmicensure}{\textbf{Output:}}
  \begin{algorithm}[htbp]
  \caption{Constructing Intrinsic Delaunay Triangulation from the Dual of Geodesic Voronoi Diagram}
  \label{alg:rdt}
  \begin{algorithmic}[1]
  \REQUIRE  $M=(V,E,F)$, a 2-manifold triangle mesh
  \ENSURE $IDT(M)$, the intrinsic Delaunay triangulation of $M$
  \STATE $GVD(V)$~=~compute\_gvd$(M)$ (Procedure \ref{alg:gvd} in Section~\ref{subsec:gvd})
  \STATE $GVD(V')$~=~ensure\_homeomorphism\_condition$(GVD(V))$, where $V\subseteq V'$ (Procedure \ref{alg:ensure_homeomorphism} in Section~\ref{subsec:homemorphism})
  \STATE $GVD(V'')$~=~ensure\_2-cell\_intersection\_condition$(GVD(V'))$, where $V'\subseteq V''$ (Procedure \ref{alg:ensure_2-cell_intersection} in Section~\ref{subsec:intersection})
  \STATE $IDT(M)$~=~compute\_dual\_graph$(GVD(V''))$ (Procedure \ref{alg:compute_dual_graph} in Section~\ref{subsec:duality})
  \end{algorithmic}
  \end{algorithm}
  }

  \ \\
  \noindent\textbf{Theorem 1 } \cite{Dyer2008}. \textit{If a geodesic Voronoi diagram satisfies the closed ball property, its dual IDT exists.
  Moreover, the IDT is regular so that 1) each geodesic triangle has three distinct \textit{g}-edges and three distinct \textit{g}-vertices;
  and 2) any two geodesic triangles, if their intersection is not empty, share either a common \textit{g}-vertex or a common \textit{g}-edge.}
  \ \\

  Dyer et al. also proved that if there are at least four distinct sites in the GVD and both the homeomorphism condition and 2-cell intersection conditions are satisfied,
  then the 3-cell intersection condition is redundant.

\section{Constructing Intrinsic Delaunay Triangulations}
\label{sec:IDT}

  This section presents the algorithm for constructing intrinsic Delaunay triangulation on meshes.
  In Sections 3.1-3.5, we assume the input 2-manifold mesh $M$ is closed.
  We then consider open meshes in Section~\ref{subsec:boundary}.
  In Section~\ref{subsec:complexity}, we analyze the time complexity of our algorithm.

  We also assume that the meshes are free of degenerate cases, i.e., no four vertices lying on a geodesic circumcircle.
  These degenerate cases can be easily handled by the symbolic perturbation technique~\cite{Edelsbrunner1990}.

  \begin{figure}[th]
  \centering
  \includegraphics[width=0.960\textwidth, bb=25 531 571 710, clip=true]{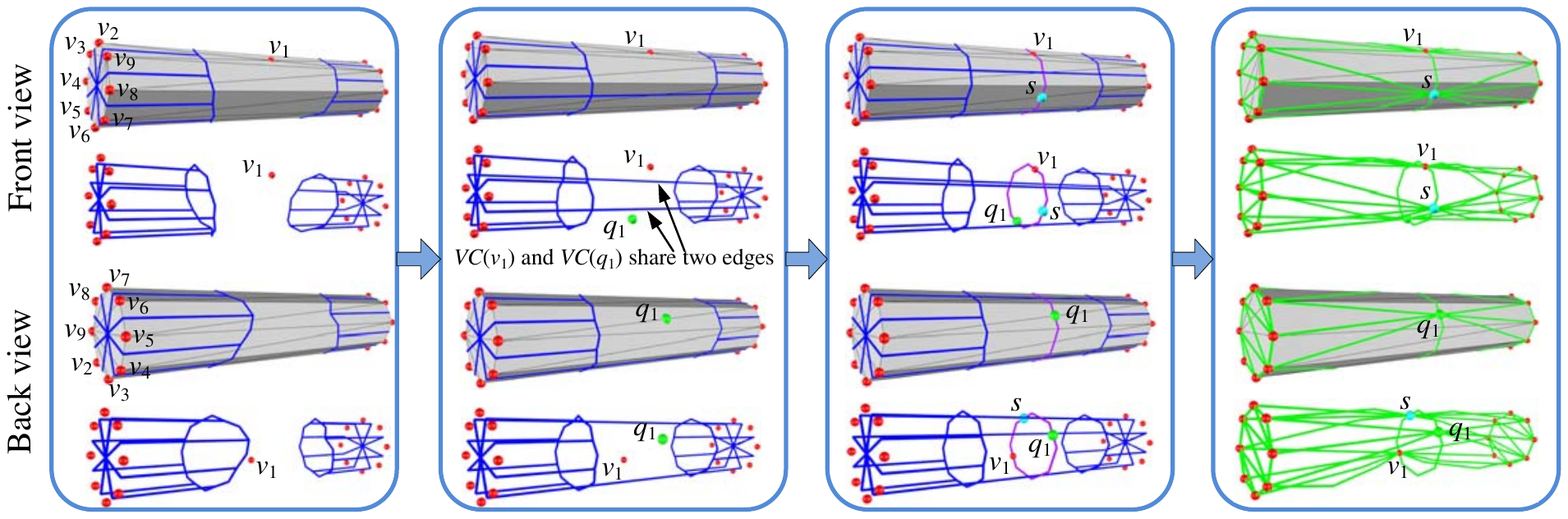}
  \begin{scriptsize}
  \makebox[1.45in]{(a) $GVD(V)$, $V=\{v_i\}_{i=1}^{17}$}
  \makebox[1.5in]{(b) $GVD(V')$, $V'=V\bigcup\{q_1\}$}
  \makebox[1.45in]{(c) $GVD(V'')$, $V''=V'\bigcup\{s\}$}
  \makebox[1.3in]{(d) Dual of $GVD(V'')$}
  \end{scriptsize}
  \vspace{-0.1in}
  \caption{Algorithmic pipeline. (a) Using the original vertices as sites, the geodesic Voronoi diagram does not have a \textit{regular} dual Delaunay triangulation,
  since the Voronoi cell $VC(v_1)$ is not simply connected.
  (b) Adding a new site $q_1$ to $V$ ensures the homeomorphism condition so that both $VC(v_1)$ and $VC(q_1)$ become topological disks.
  However, they share two common edges (see the front and back views).
  (c) Adding another site $s$, which is equidistant to $v_1$ and $q_1$, the new geodesic Voronoi diagram $GVD(V'')$ satisfies the closed ball property.
  (d) The dual graph of $GVD(V'')$ is a regular Delaunay triangulation.
  Color schemes: the mesh vertices are red dots and the added auxiliary sites are green.
  The mesh edges are grey, the Voronoi edges blue, the Delaunay edges green, and the geodesics purple, respectively.}
  \label{fig:pipeline}
  \end{figure}

\subsection{Overview}
\label{subsec:overview}

  Flipping edges and computing the dual of Voronoi diagrams are two commonly used techniques for constructing Delaunay triangulations in $\mathbb{R}^2$.
  As mentioned before, the edge-flip algorithm on meshes~\cite{Indermitte2001,Fisher2006} does not have a known time complexity and it may produce non-regular triangulations,
  which have self-loops or faces with only two edges.
  In this paper, we take the other direction by computing the dual of geodesic Voronoi diagrams.
  The major challenge in this direction is that not every GVD has a dual Delaunay triangulation.
  As shown in Section~\ref{subsec:closedball}, the closed ball property is the sufficient condition for the existence of a dual Delaunay triangulation.
  Therefore, our goal is to enforce the closed ball property everywhere on the GVD.
  We consider only the homeomorphism condition and the 2-cell intersection condition in our algorithm, since all the models we are dealing with have more than 4 vertices.

  Our algorithm (ref. Algorithm \ref{alg:rdt}) consists of four steps.
  Firstly, taking all mesh vertices as sites, it computes the geodesic Voronoi diagram $GVD(V)$.
  Secondly, it checks the homeomorphism condition for all Voronoi cells. If a Voronoi cell, say $VC(v_i)$, is not homeomorphic to a disk,
  the algorithm adds an auxiliary site in $VC(v_i)$ and then locally updates the GVD.
  Thirdly, it checks the 2-cell intersection condition for all pairs of adjacent Voronoi cells.
  If two adjacent Voronoi cells, say $VC(v_i)$ and $VC(v_j)$, have more than 1 common Voronoi edges,
  the algorithm adds an auxiliary site that is equidistant to $v_i$ and $v_j$, and locally updates the GVD.
  After steps 2 and 3, the updated GVD is guaranteed to have a dual triangulation, which is computed in Step 4.

  Intuitively speaking, our method adaptively increases the sampling density for the regions where the closed ball property fails.
  Figure~\ref{fig:pipeline} illustrates the algorithmic pipeline using a toy cylinder model.

\subsection{Computing the GVD}
\label{subsec:gvd}

  The first step in our algorithm is to construct the geodesic Voronoi diagram on $M$.
  Liu et al.~\cite{Liu2011} presented a generic algorithm, which runs in $O(n^2\log n)$ time for $m (\leq n)$ sites, where $n=|V|$ is the number of vertices in $M$.
  One, of course, can apply Liu et al.'s algorithm directly to our application.
  However, our scenario is slightly different in that we take \textit{all} mesh vertices as sites.
  Moreover, as our goal is to compute the dual Delaunay triangulation, we don't need to \textit{explicitly} construct the Voronoi diagram.
  Therefore, we adapt Liu et al's algorithm for IDT construction (see Procedure~\ref{alg:gvd}).
  We will show in Section~\ref{subsec:complexity} that our adapted GVD algorithm runs \textit{empirically} in linear time on real-world models.

  {
  \renewcommand{\algorithmicensure}{\textbf{Postcondition:}}
  \floatname{algorithm}{Procedure}
  \begin{algorithm}[h]
  \caption{compute\_gvd$(M)$}
  \label{alg:gvd}
  \begin{algorithmic}[1]
  \STATE Taking $\{v_1,v_2,\cdots,v_n\}$ as sources, compute the geodesic distance by the MMP algorithm~\cite{Mitchell1987}.
  \STATE Identify the triangles at which three or more bisectors meet.
  \STATE Compute the \textit{symbolic} representation of Voronoi edges and Voronoi cells.
  \ENSURE The symbolic representation of $GVD(V)$
  \end{algorithmic}
  \end{algorithm}
  }

  Using the mesh vertices as sources, we compute the geodesic distances by the Mitchell-Mount-Papadimitriou (MMP) algorithm~\cite{Mitchell87}.
  Each mesh vertex is assigned 0 (since it is the source) and each mesh edge is partitioned into disjoint intervals, called \textit{windows},
  where a window encodes the shortest distance to some source vertex.
  Edge $e$ contains a bisector if $e$ has windows corresponding to different sources.
  Then, we identify the triangles where three or more bisectors meet.
  Rather than computing the \textit{geometry} of each Voronoi cell explicitly,
  we need only a symbolic representation:
  the bisector of $v_i$ and $v_j$ is an \textit{ordered} pair $\{v_i,v_j\}$ with $i<j$,
  and a Voronoi cell is an ordered list $VC(v_i)=\{v_{i_1},v_{i_2},\cdots\}$, where each pair of $v_i$ and $v_{i_k}$ corresponds to a trimmed bisector, i.e., a Voronoi edge.

  In case the geometry of a Voronoi cell is required, which is rarely encountered in IDT construction on real-world models,
  the bisectors can be computed on-the-fly.
  To compute the boundary edge $b(v_{i},v_{i_{k}})$ of $VC(v_i)$,
  we find the triangles containing the two branch points $\{v_i,v_{i_{k-1}},v_{i_k}\}$ and $\{v_i,v_{i_k},v_{i_{k+1}}\}$,
  then we locally trace the bisector by unfolding the corresponding triangles onto $\mathbb{R}^2$.

\subsection{Ensuring the Homeomorphism Condition}
\label{subsec:homemorphism}

  In $\mathbb{R}^2$, the term ``bisector'' refers to the set of points which are equidistant to two \textit{distinct} sites.
  In contrast to the 2D counterpart, there are two types of bisectors on a mesh:
  one is the conventional bisector of two distinct sites, and the other is a special bisector, called \textit{pseudo bisector}, which corresponds to a \textit{single} site.

  \ \\
  \noindent\textbf{Definition 5} (Pseudo-bisector). \textit{The pseudo-bisector of a site $v$ consists of points $q\in VC(v)$ such that there are two or more geodesic paths between $v$ and $q$ of equal length.}
  \ \\

  Pseudo bisectors usually occur in a cylinder-shaped region.
  See Figure~\ref{fig:pipeline}(a).
  The following proposition states that a pseudo bisector, if exists, must be a line segment after unfolding.

  \ \\
  \noindent\textbf{Proposition 1.}~\textit{Let $GVD(P)$ be the geodesic Voronoi diagram of sites $P$, where $V\subseteq P$.
  If, for a site $p\in P$, the Voronoi cell $VC(p)$, has a pseudo bisector $pb(p)$, then $pb(p)$ is a line segment when making all faces containing $pb(p)$ coplanar.}
  \ \\

  Since pseudo bisectors are due to low sampling density, we add an auxiliary site on each pseudo-bisector to destroy it.
  Given a Voronoi cell $VC(p_i)$ which contains a pseudo bisector $pb(p_i)$, we compute a point $q\in pb(p_i)$ such that $q$ minimizes the geodesic distance $d(p_i,x)$,
  $\forall x\in pb(p_i)$.
  If the unfolded images of $\gamma(p_i,q)$ and $pb(p_i)$ are perpendicular, $q$ is the intersection point.
  Otherwise, $q$ is one of $pb(p_i)$'s two end points.
  We then locally update the GVD around the new site $q$.
  The following proposition guarantees that Voronoi cell $VC(p_i)$ becomes a topological disk when the auxiliary site $q$ is added.

  \ \\
  \noindent\textbf{Proposition 2}. \textit{Assume that Voronoi cell $VC(p_i)$ has a pseudo-bisector $pb(p_i)$ and the point $q\in pb(p_i)$ minimizes the geodesic distance $d(p_i,x)$ for all $x\in pb(p_i)$.
  Add the auxiliary site $q$ into the GVD.
  Then both $VC(q)$ and $VC(p_i)$ are topological disks.}
  \ \\

{
\renewcommand{\algorithmicensure}{\textbf{Postcondition:}}
\floatname{algorithm}{Procedure}
\begin{algorithm}[htb]
\caption{ensure\_homeomorphism\_condition$(GVD(V))$}
\label{alg:ensure_homeomorphism}
\begin{algorithmic}[1]
\STATE $V' = V$
\FOR {every $v_i\in V$}
\IF {$VC(v_i)$ has a pseudo-bisector $pb(v_i)$}
\STATE Compute $q\in pb(v_i)$ that minimizes $d(v_i,x)$, $x\in pb(v_i)$
\STATE $V'=V\bigcup\{q\}$
\STATE Locally update $GVD(V')$
\ENDIF
\ENDFOR
\ENSURE Each Voronoi cell in $GVD(V')$ is homeomorphic to a disk
\end{algorithmic}
\end{algorithm}
}

\ \\
\noindent\textbf{Proposition 3}. \textit{$GVD(V')$ has at most $4n+4g-4$ Voronoi vertices, $6n+6g-6$ Voronoi edges, and $2n$ Voronoi cells, where $n=|V|$ and $g$ is the genus of $M$.}
\ \\

  \begin{figure}[th]
  \centering
  \includegraphics[width=0.55\textwidth, bb=15 448 578 680, clip=true]{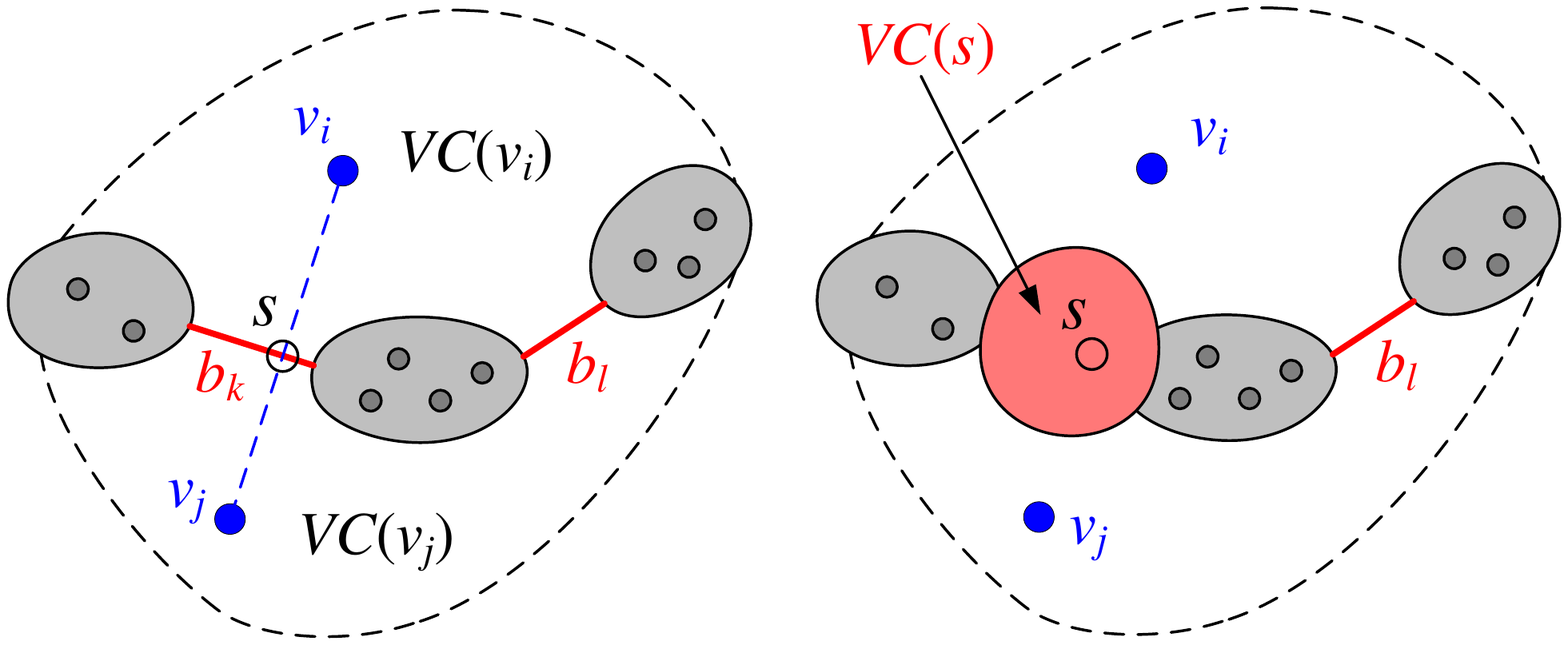}
  \makebox[1.6in]{(a)}\makebox[1.6in]{(b)}\\
  \vspace{-0.1in}
  \caption{Voronoi cells $VC(v_i)$ and $VC(v_j)$ violate the 2-cell intersection condition, since they share two Voronoi edges $b_k$ and $b_l$.
  To destroy one Voronoi edge, say $b_k$, we add an auxiliary site $s\in b_k$ and update the GVD locally.
  The gray shaded areas are some Voronoi cells, which are irrelevant to the discussion.}
  \label{fig:condition2}
  \end{figure}

\subsection{Ensuring the 2-cell Intersection Condition}
\label{subsec:intersection}

  After adding auxiliary sites on the pseudo-bisectors, all Voronoi cells in $GVD(V')$ are simply connected.
  Let $B$ denote the set of Voronoi edges (i.e., trimmed bisectors).
  Recall that the symbolic representation of the bisector of $v_i$ and $v_j$ is an ordered pair $\{v_i,v_j\}$ satisfying $i<j$.
  We sort all the Voronoi edges $\{v_i,v_j\}$ in $B$ in the ascending order by their first elements.
  If two ordered pairs have the same first element, the second element is used to break a tie.

  If an order pair $\{v_i,v_j\}$ appears more than once in the sorted list $B$,
  the two corresponding Voronoi cells $VC(v_i)$ and $VC(v_j)$ violate the 2-cell intersection condition.
  Let $b_k$ and $b_l$ be the two common edges shared by $VC(v_i)$ and $VC(v_j)$. See Figure \ref{fig:condition2}(a).
  To destroy one edge, say, $b_k$, we compute a point $s\in b_k$ which minimizes the geodesic distance $d(p_i,x)$, $\forall x\in b_k$.
  As $b_k$ bisects $p_i$ and $p_j$, the path $\gamma(p_j,s)$ is a minimizing geodesic.
  We then add $s$ into $V'$ and update the Voronoi cells locally. See Figure \ref{fig:condition2}(b).
  The following proposition guarantees that the updated Voronoi cells satisfy both the homeomorphism condition and the 2-cell intersection condition.

  \ \\
  \noindent\textbf{Proposition 4}. \textit{Let $VC(v_i)$ and $VC(v_j)$ be adjacent Voronoi cells that share two Voronoi edges $b_k$ and $b_l$.
  The point $s\in b_k$ minimizes $d(v_i,x)$ for $x\in b_k$.
  Then the GVD with additional site $s$ satisfies both the homeomorphic condition and the 2-cell intersection condition.}
  \ \\

  {
  \renewcommand{\algorithmicensure}{\textbf{Postcondition:}}
  \floatname{algorithm}{Procedure}
  \begin{algorithm}[htb]
  \caption{ensure\_2-cell\_intersection\_condition$(GVD(V'))$}
  \label{alg:ensure_2-cell_intersection}
  \begin{algorithmic}[1]
  \STATE Let $V'' = V'$
  \STATE Sort all Voronoi edges in the ascending order
  \FOR {any two Voronoi edges $b_k=\{p_i,p_j\}$ and $b_l=\{p_i,p_j\}$ with the same ordered pair}
  \STATE Compute point $s\in b_k$ that minimizes $d(p_i,x)$, $x\in b_k$
  \STATE $V''=V''\bigcup\{s\}$
  \STATE Locally update $GVD(V'')$
  \ENDFOR
  \ENSURE $GVD(V'')$ satisfies the closed ball property
  \end{algorithmic}
  \end{algorithm}
  }

  \ \\
  \noindent\textbf{Proposition 5}. \textit{$GVD(V'')$ has $O(n)$ Voronoi vertices, Voronoi edges and Voronoi cells.}
  \ \\

  \begin{figure}[th]
  \centering
  \includegraphics[width=2.0in, bb=61 292 530 668, clip=true]{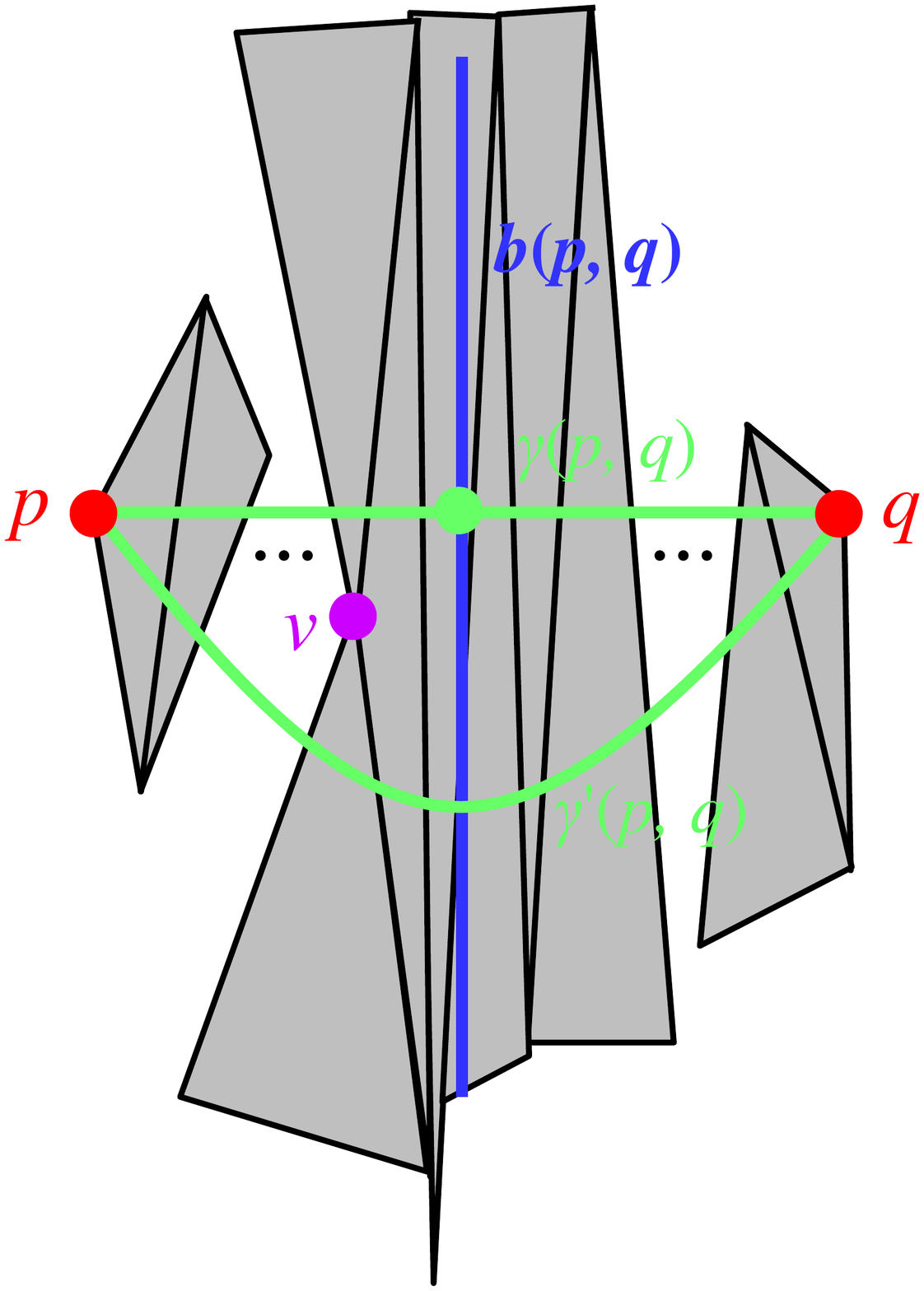}
  \vspace{-0.1in}
  \caption{Each Voronoi edge corresponds to a \textit{unique} Delaunay edge.
  Consider two \textit{adjacent} Voronoi cells $VC(p)$ and $VC(q)$. Let $\Omega\triangleq VC(p)\bigcup VC(q)$.
  Assume there are two geodesic paths $\gamma(p,q)\in \Omega$ and $\gamma'(p,q)\in \Omega$, which connect $p$ and $q$.
  Let $\mathcal{F}$ (resp. $\mathcal{F'}$) denote the set of faces that $\gamma(p,q)$ (resp. $\gamma'(p,q)$) passes through.
  Since $\gamma(p,q)\neq \gamma'(p,q)$, the two face sets are different $\mathcal{F}\neq \mathcal{F'}$,
  meaning that there is at least one vertex, say $v$, inside the region bounded by $\gamma(p,q)$ and $\gamma'(p,q)$.
  As a result, the Voronoi cell $VC(v)$ is in between $VC(p)$ and $VC(q)$, which is a contradiction.
  Therefore, there is a unique geodesic path $\gamma(p,q)\in\Omega$ between $p$ and $q$, which is taken as the Delaunay edge dual to $b(p,q)$. }
  \label{fig:bisector-geodesic}
  \end{figure}

\subsection{Computing the Dual Graph}
\label{subsec:duality}

  Since the closed ball property holds everywhere for $GVD(V'')$, the dual Delaunay triangulation exists.
  By the 2-cell intersection condition, two adjacent Voronoi cells $VC(p)$ and $VC(q)$ share only one common Voronoi edge, denoted by $b(p,q)$.
  As Figure~\ref{fig:bisector-geodesic} shows, the Voronoi edge $b(p,q)$ corresponds to a \textit{unique} Delaunay edge, which is dual to $b(p,q)$.

  Given a Voronoi edge $b(p,q)$ in $GVD(V'')$, we compute its dual Delaunay edge by using the geodesic distance field obtained in Procedure 2 (line 1).
  Note that the MMP algorithm splits each mesh edge into two oriented halfedges with opposite directions
  and each halfedge contains the windows, a discrete data structure that encodes the geodesic paths to the closest source.
  Since both $VC(p)$ and $VC(q)$ do not contain any mesh vertex other than the generating sites $p$ and $q$,
  each side of $b(p,q)$ contains exactly one window, which encodes the geodesic paths to $p$ and $q$, respectively.
  We denote by $\mathcal{F}_p$ (resp. $\mathcal{F}_q$) the set of faces containing the geodesic paths from $p$ (resp. $q$) to any point on $b(p,q)$.
  Then the unique Delaunay edge $\gamma(p,q)$ is computed by unfolding the faces $\mathcal{F}_p\cup\mathcal{F}_q$.

  {
  \renewcommand{\algorithmicensure}{\textbf{Postcondition:}}
  \floatname{algorithm}{Procedure}
  \begin{algorithm}[htb]
  \caption{compute\_dual\_graph$(GVD(V''))$}
  \label{alg:compute_dual_graph}
  \begin{algorithmic}[1]
  \FOR {each Voronoi edge $b(p,q)\in B$ in $GVD(V'')$}
  \STATE Using the windows stored at the two sides of $b(p,q)$, compute the face sets $\mathcal{F}_p$ and $\mathcal{F}_q$, respectively.
  \STATE Compute $\gamma(p,q)$ by unfolding the triangles in $\mathcal{F}_p\bigcup\mathcal{F}_q$.
  \ENDFOR
  \FOR {any three Voronoi cells that meet at a Voronoi vertex}
  \STATE Create a geodesic triangle with the three corresponding \textit{g}-edges.
  \ENDFOR
  \ENSURE $IDT(M)$ is a regular IDT.
  \end{algorithmic}
  \end{algorithm}
  }

  \begin{figure}[h]
  \centering
  \includegraphics[width=3.0in, bb=106 590 413 654, clip=true]{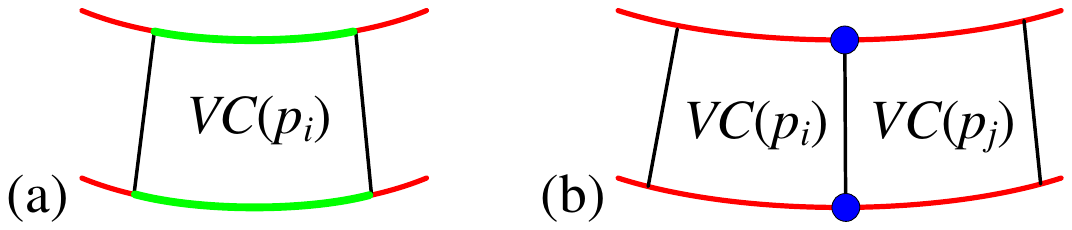}
  \vspace{-0.1in}
  \caption{The closed ball property for manifolds with boundary. (a) $VC(p_i)\cap \partial M$ has two non-adjacent boundary Voronoi edges (in green), which violate the condition A1. Case (b) $VC(p_i)\cap VC(p_j)\cap\partial M$ has two boundary Voronoi vertices (blue dots), which violate the condition A2. The boundary $\partial M$ is shown in red.}
  \label{fig:manifold-bndy}
  \end{figure}

\subsection{Manifolds with Boundaries}
\label{subsec:boundary}

  Now we consider meshes with boundaries. Let us denote by $\partial M$ the boundary of $M$.
  The closed ball property for a manifold with boundary has two more conditions \cite{Edelsbrunner1997}:
  \begin{itemize}
  \item[A1.] The intersection of any single Voronoi cell and $\partial M$ is empty, a single point or a single line segment homeomorphic to $[0,1]$.
  \item[A2.] The intersection of two distinct Voronoi cells and $\partial M$,  $VC(p_i)\bigcap VC(p_j)\bigcap\partial M$, $i\neq j$, is either empty or a single point.
  \end{itemize}

  These additional conditions complement to the three conditions of the closed ball property for closed manifolds.
  Figure \ref{fig:manifold-bndy} shows two cases where either condition A1 or A2 does not hold.
  These issues can be fixed by adding auxiliary sites to the boundary $\partial M$.

  \ \\
  \noindent\textbf{Proposition 6}. \textit{The Voronoi cell $VC(v)$ in $GVD(V'')$ has two non-adjacent boundary Voronoi edges $b_k$ and $b_l$
  (i.e., $b_k,b_l\in\partial M$ and $b_k\cap b_l=\emptyset$).
  Without loss of generality, say $v\notin b_k$.
  The point $s\in b_k$ minimizes the geodesic distance $d(v,x)$ for $x\in b_k$.
  The GVD with the additional site $s$ satisfies both boundary conditions A1 and A2.}
  \ \\

  {
  \renewcommand{\algorithmicensure}{\textbf{Postcondition:}}
  \floatname{algorithm}{Procedure}
  \begin{algorithm}[htb]
  \caption{ensure\_boundary\_condition$(GVD(V''))$}
  \label{alg:ensure_boundary_condition}
  \begin{algorithmic}[1]
  \STATE Let $V''' = V''$
  \IF {a Voronoi cell $VC(v)$ has two non-adjacent boundary Voronoi edges}
  \STATE Set $b_k$ be the boundary Voronoi edge that does not contain $v$
  \STATE Compute point $s\in b_k$ that minimizes $d(v,x)$, $x\in b_k$
  \STATE $V'''=V'''\bigcup\{s\}$
  \STATE Locally update $GVD(V''')$
  \ENDIF
  \ENSURE $GVD(V''')$ satisfies the boundary conditions A1 and A2
  \end{algorithmic}
  \end{algorithm}
  }

  \ \\
  \noindent\textbf{Proposition 7}. \textit{$GVD(V''')$ has $O(n)$ Voronoi vertices, Voronoi edges and Voronoi cells.}
  \ \\

\subsection{Complexity Analysis}
\label{subsec:complexity}

  \noindent\textbf{Theorem 2}. \textit{Given a closed 2-manifold mesh $M=(V,E,F)$,
  our IDT construction algorithm (Algorithm 1) has a worst-case time complexity $O(n^2\log n)$, where $n=|V|$.}

  \begin{figure}[th]
  \centering
  \includegraphics[width=2.6in, bb=43 443 514 681, clip=true]{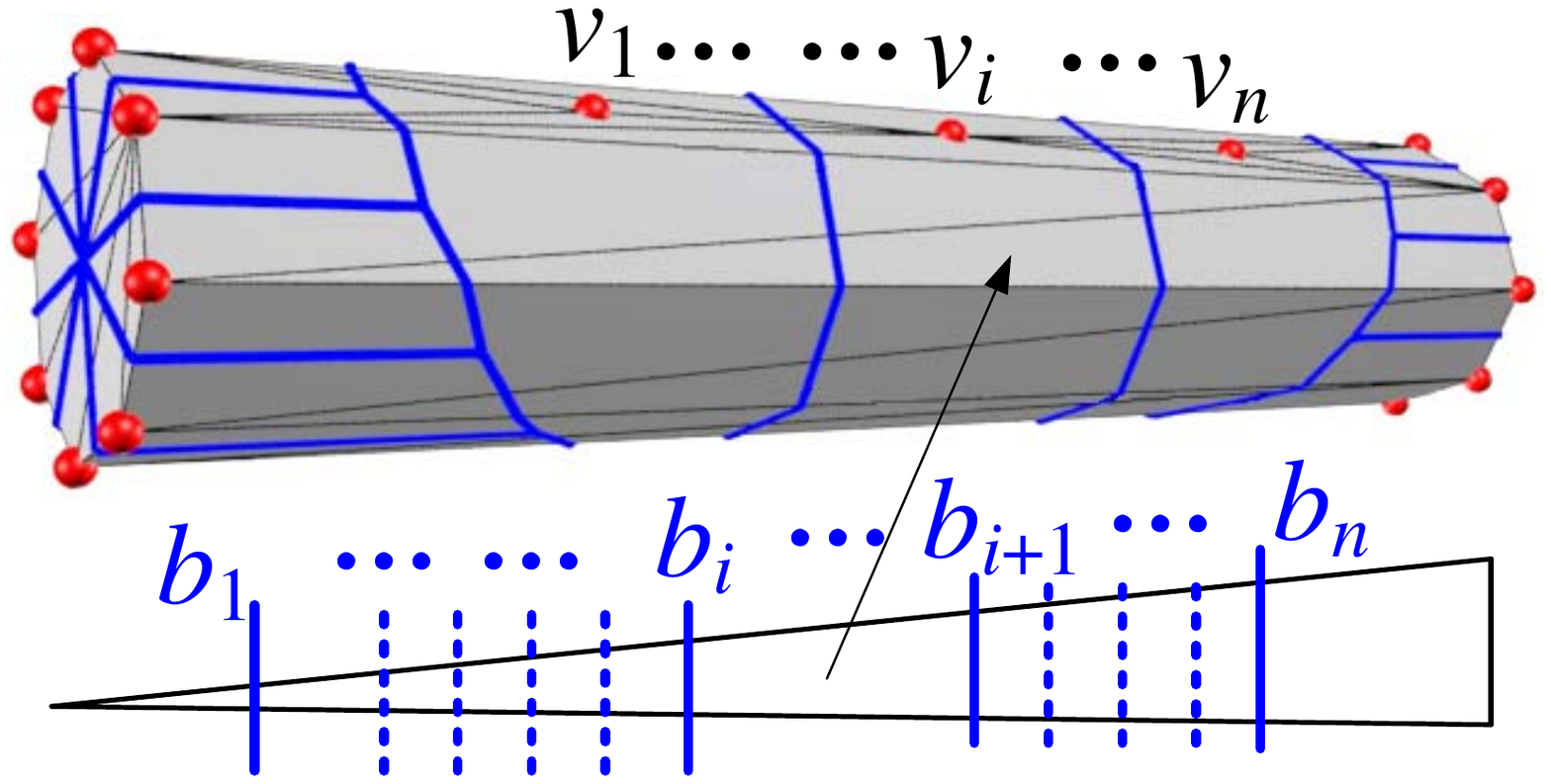}
  \\\makebox[2.5in]{(a) Worst case} \\
  \includegraphics[width=1.5in]{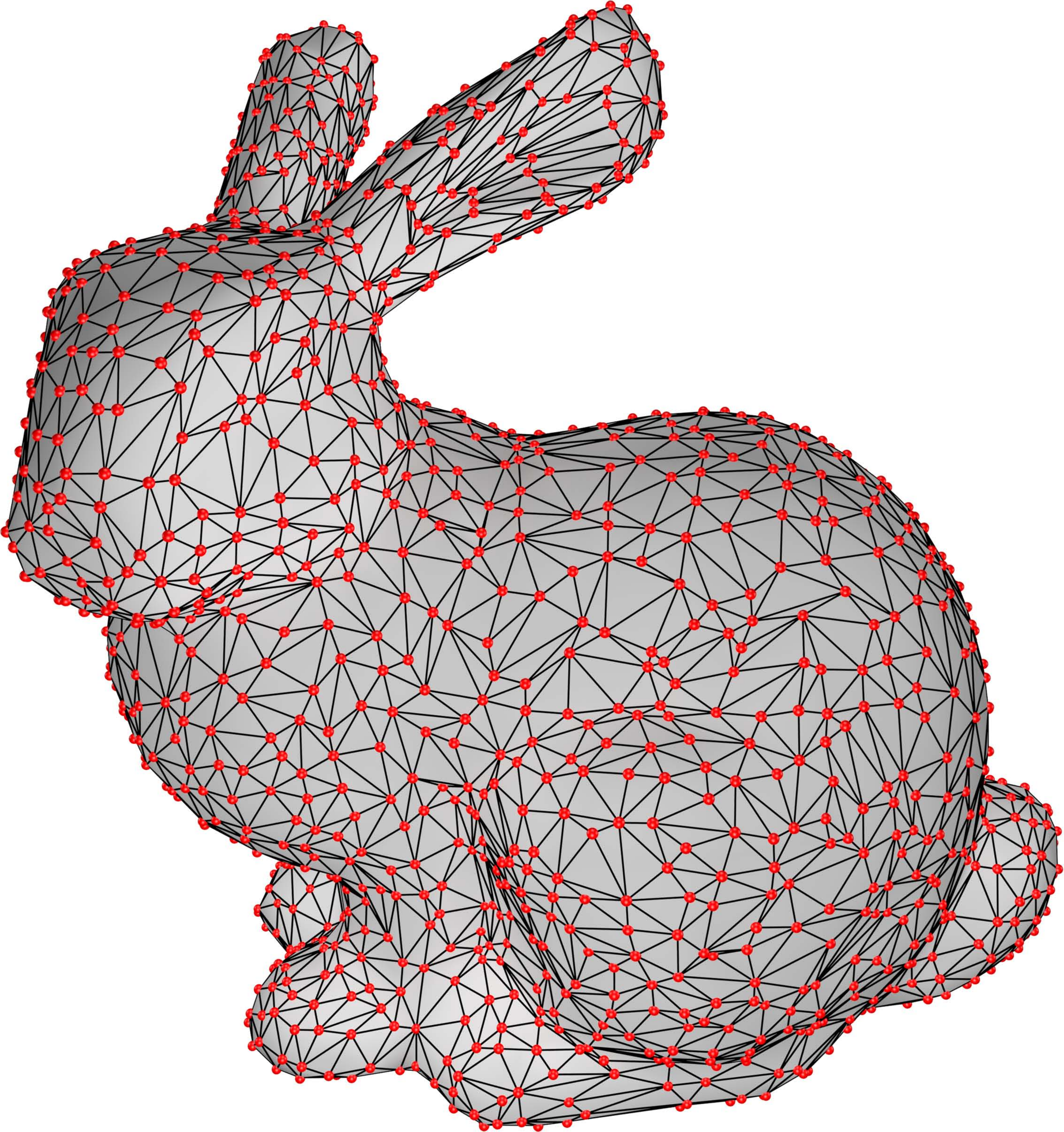}
  \includegraphics[width=1.5in]{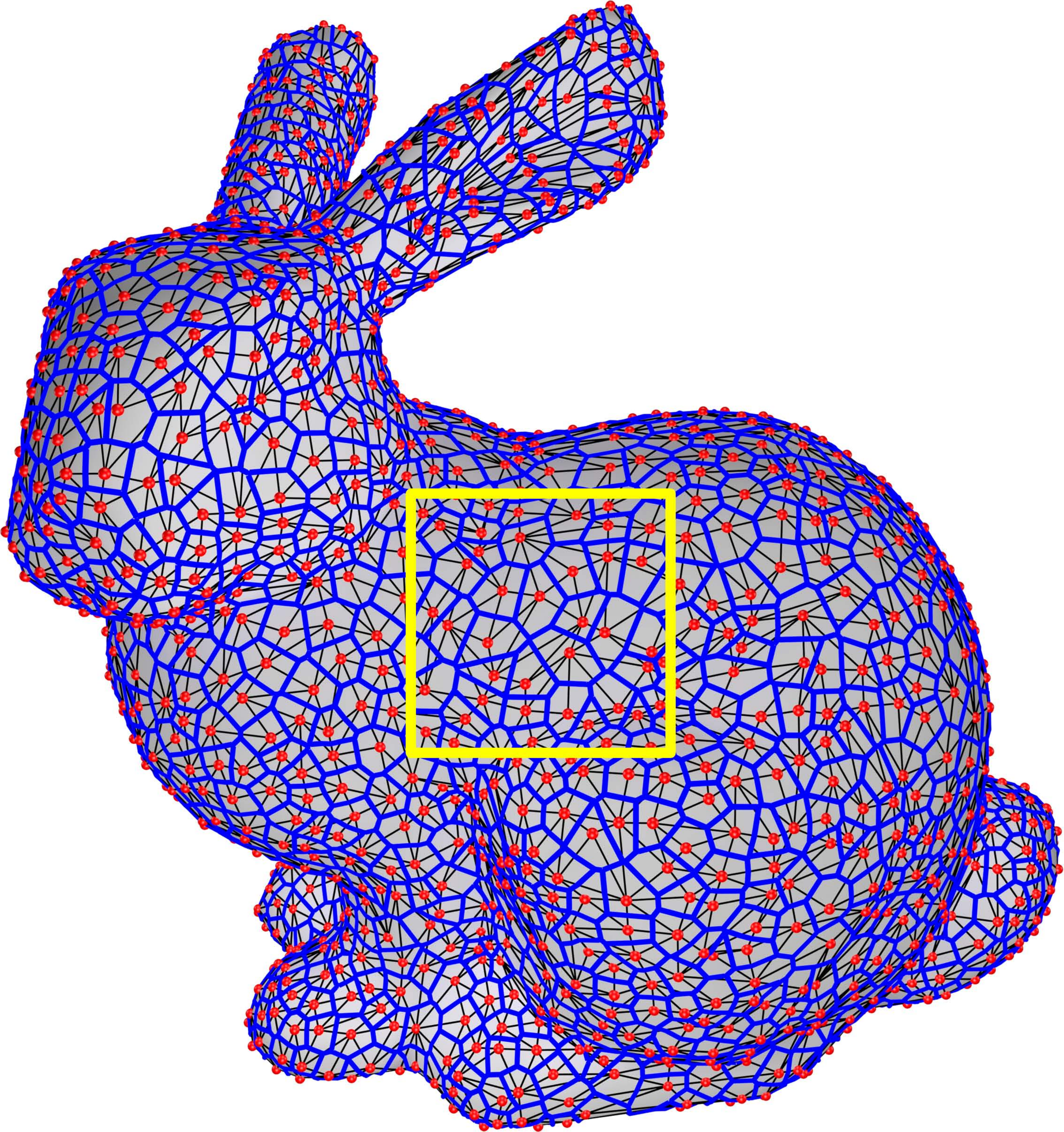}
  \includegraphics[width=1.5in]{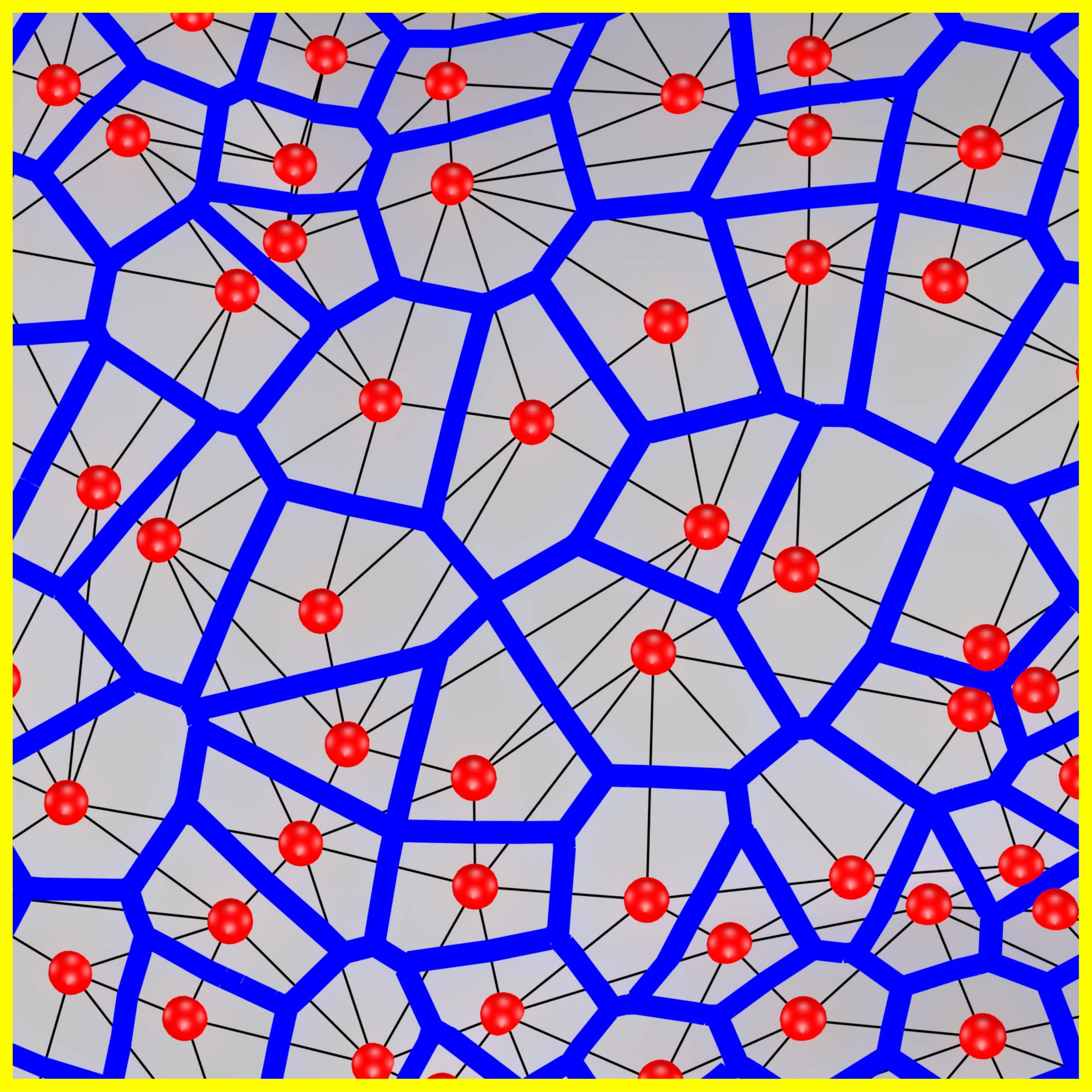}\\
  \makebox[3.1in]{(b) Normal case}\\
  \vspace{-0.1in}
  \caption{Complexity of bisectors. (a) The triangles on the side of the cylinder are \textit{global}, since they span the whole model.
  Each of these global triangle contains $O(n)$ bisectors.
  As a result, computing $GVD(V)$ takes $O(n^2\log n)$ time.
  (b) In general, if the majority of mesh faces are \textit{local} (i.e., each triangle covers only a local region on $M$),
  most edges intersect only 1 or 2 bisectors (see the close-up view).
  Therefore, the GVD can be computed in $O(n)$ time.
  }
  \label{fig:bisector-worst-case}
  \end{figure}

\ \\
  \noindent\textit{Proof.} Our algorithm consists of four steps, corresponding to Procedures 2, 3, 4 and 5.

  First, we show that Step 1, computing the geodesic Voronoi diagram $GVD(V)$, takes $O(n^2\log n)$ time, which is the dominant term of all four steps.
  The GVD algorithm is built upon the MMP algorithm~\cite{Mitchell1987}, a classic discrete geodesic algorithm which represents the discrete geodesic wavefront using windows.
  Maintaining the windows in a priority queue, the MMP algorithm iteratively propagates windows across the faces and updates the geodesic distance when a window covers a vertex or part of an edge.
  It terminates when the priority queue is empty, i.e., all vertices and edges have been covered by some windows.
  Mitchell et al. showed that each edge has at most $O(n)$ windows, thus, the MMP algorithm has a worst-case time complexity $O(n^2\log n)$.
  In the GVD computation, $O(n)$ windows on an edge $e$ means $e$ has $O(n)$ bisectors.
  Then lines 2 and 3 in Procedure 2 have $O(n^2)$ time complexity.
  Therefore, Step 1 takes $O(n^2\log n)$ time.

  Then we show that Steps 2, 3 and 4 take $O(n^2)$ time.
  Given a Voronoi cell $VC(p_i)$ that contains a pseudo-bisector $pb(p_i)$, finding the point $q$ that minimizes $d(p_i,x)$, $\forall x\in pb(p_i)$, takes O(1) time.
  Locally updating $VC(q)$ takes at most $O(n)$ time.
  By Proposition 3, there are at most $n$ auxiliary sites added in Step 2.
  Thus, Step 2 takes $O(n^2)$ time.

  Similarly, by Proposition 5, there are $O(n)$ auxiliary sites added in Step 3,
  and updating the GVD for each new site takes $O(n)$ time. Therefore, Step 3 takes $O(n^2)$ time.

  In Step 4, computing a Delaunay edge $\gamma(p,q)$ takes $O(n)$ time, since there are at most $O(n)$ faces in the set $\mathcal{F}_p\bigcup\mathcal{F}_q$.
  As $GVD(V'')$ has $O(n)$ Voronoi edges, Step 4 also takes $O(n^2)$ time.

\begin{figure}[htbp]
  \centering
  \includegraphics[width=0.66\textwidth, bb=33 429 573 736, clip=true]{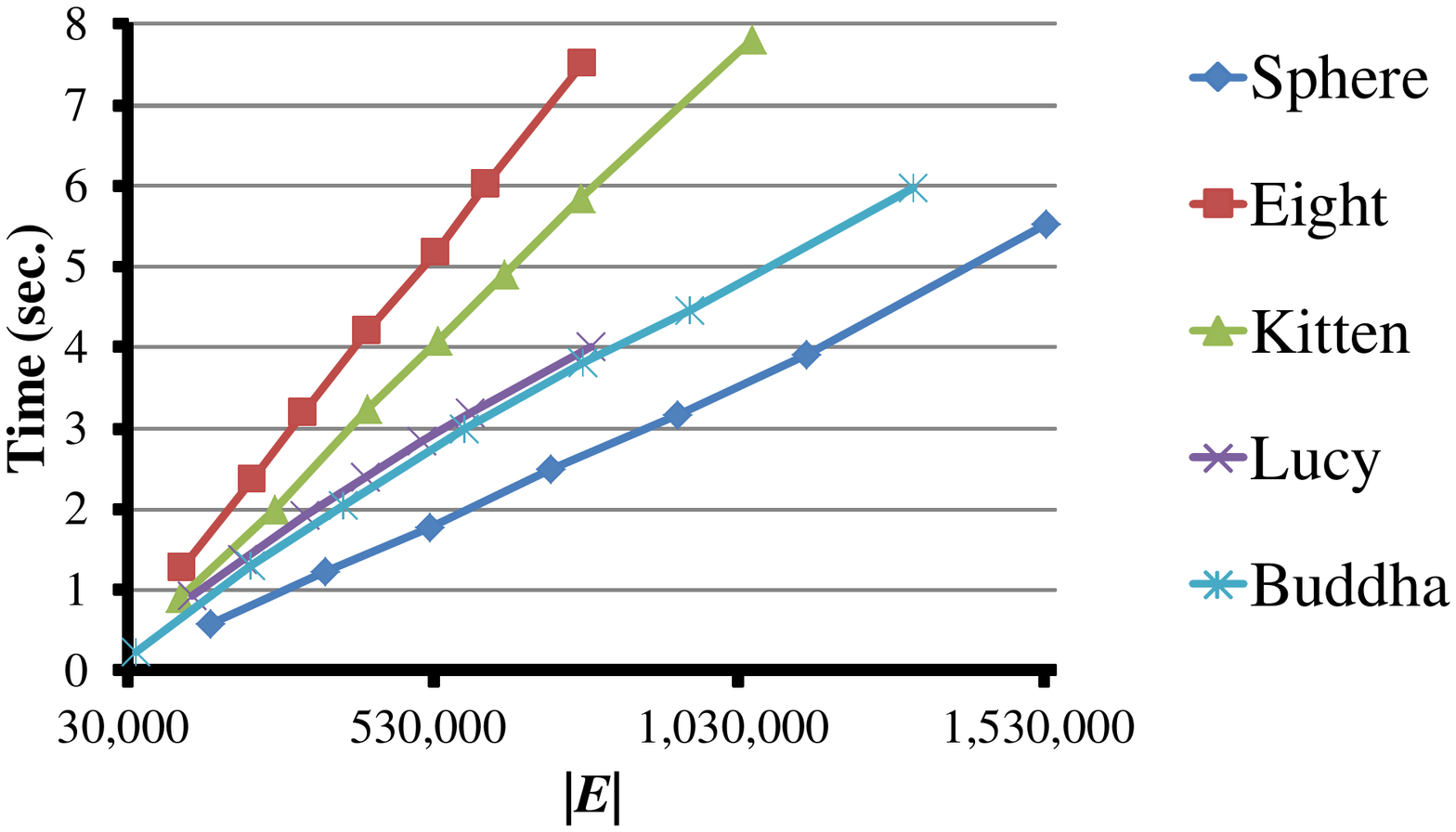}
  \vspace{-0.1in}
  \caption{Our algorithm runs in linear time empirically. The horizontal axis shows the mesh complexity and the vertical axis is the execution time (in seconds).}
  \label{fig:empirical-time}
  \end{figure}

  Putting it all together, our algorithm has a worst-case time complexity $O(n^2\log n)$.
~~~~~~~~~~~~~~~~~~~~~~~~~~~~~~~~~~~~~~~~~~~~~~~~~~~~~~~~~~~~~~~~~~\qed

  \ \\
  The \textit{theoretical} worst-case time complexity $O(n^2\log n)$ is very pessimistic,
  since it happens only when each triangle contains $O(n)$ bisectors.
  See Figure~\ref{fig:bisector-worst-case}(a) for a model with the worst-case time complexity.
  We call a triangle with $O(n)$ bisectors \textit{global}, since it indeed spans a global region on the model.
  We observe that on many real-world models, the majority of the triangles are not global,
  even though the mesh triangulation is poor (i.e., far from its Delaunay triangulation).
  Computational results show that on average, a mesh edge on a real-world model has only $O(1)$ bisectors.
  See Figure~\ref{fig:bisector-worst-case}(b).
  As a result, all of the four steps in our algorithm run in $O(n)$ time.
  Computational results in Figure~\ref{fig:empirical-time} confirm that our algorithm has an empirical linear time complexity.

  \begin{table}
  \begin{center}
  \begin{small}
  \begin{tabular}{|l|c|cc|cc|cc|}
  \hline
  \multicolumn{1}{|l|}{\multirow {2}{*}{Model ($|E|$)}} & \multicolumn{1}{c|}{\multirow{2}{*}{$(\sigma_{mean},\sigma_{max},\sigma_{std})$}} & \multicolumn{2}{c|}{Edge-flipping} & \multicolumn{2}{c|}{Ours} & \multicolumn{2}{c|}{$\mathrm{cond}(\bigtriangleup)$}\\
  \cline{3-8}
       &     & Time & $N_{fe}$ & Time & $N_{as}$ & $M$ & $IDT(M)$\\
  \hline
  \hline
  CSG (357)      & (4.74,33.02,2.13)   & 0.006 & 48.4\%   & 0.004 & 1 & $1.3\times 10^{5}$ & $4.9\times 10^{3}$\\
  Fandisk (1K)      & (44.2, 177.5, 9.49) & 0.015 & 37.3\%   & 0.011 & 0 & $2.4\times 10^7$ & $7.7\times 10^{4}$ \\
  Eight (3K)       & (1.67, 11.80, 0.88) & 0.031 &  23.6\%  & 0.016 & 0 & $7.6\times 10^4$ & $3.9\times10^{4}$ \\
  Sphere (6K)       & (1.67, 59.60, 2.14) & 0.019 &  0.08\%  & 0.006 & 0 & $5.5\times 10^5$ & $5.5\times 10^5$ \\
  Teapot (17K)      & (1.69, 20.74, 1.66) & 0.171 &  23.5\%  & 0.125 & 0 & $8.7\times 10^5$ & $3.4\times 10^5$ \\
  Decocube (24K)    & (1.56, 13.43, 0.65) & 0.188 &  17.6\%  & 0.140 & 0 & $5.3\times 10^5$ & $2.5\times 10^5$ \\
  Fertility (37K)   & (2.34, 17.20, 1.11) & 0.609 &  46.3\%  & 0.561 & 0 & $8.5\times 10^8$ & $2.0\times 10^8$ \\
  Crank (60K) & (17.5,279.5,14.2) & 1.539 & 47.6\% & 1.350 & 0 & $1.9\times 10^{10}$ & $9.5\times 10^6$\\
  Bunny (216K) & (1.23, 11.83, 0.19) & 0.780 &  1.40\%  & 0.156 & 0 & $4.1\times 10^6$ & $2.6\times 10^6$ \\
  Armadillo (519K) & (1.31, 95.29, 2.39) & 2.324 &  4.59\%  & 0.983 & 0 & $1.1\times 10^7$ & $3.2\times 10^6$ \\
  Lucy (789K) & (1.45, 34.24, 1.38) & 5.179 &  9.44\%  & 3.960 & 0 & $5.8\times 10^7$ & $4.3\times 10^7$ \\
  Buddha (1,196K) & (1.47, 29.16, 5.22) & 8.502 &  8.66\%  & 5.379 & 0 & $6.8\times 10^7$ & $3.6\times 10^7$ \\
  \hline
  \end{tabular}
  \end{small}
  \end{center}
  \caption{Mesh complexity and performance statistics. $|E|$: the number of edges in $M$;
  $N_{fe}$ (\%): percentage of the edges that are flipped by the edge-flipping algorithm; $N_{as}$: the number of auxiliary sites added by our algorithm;
  $\mathrm{cond}(\bigtriangleup)$: the condition number of the discrete Laplacian matrix.
  The 3-tuple $(\sigma_{mean},\sigma_{max},\sigma_{std})$ shows the mean, max and standard deviation of the anisotropy measure $\sigma$.
  The running time was measured on an Intel Core i7-2600 CPU (3.40 GHz).}
  \label{table:runningtime}
  \end{table}

  \begin{figure}[htbp]
  \centering
  \includegraphics[width=3.8in]{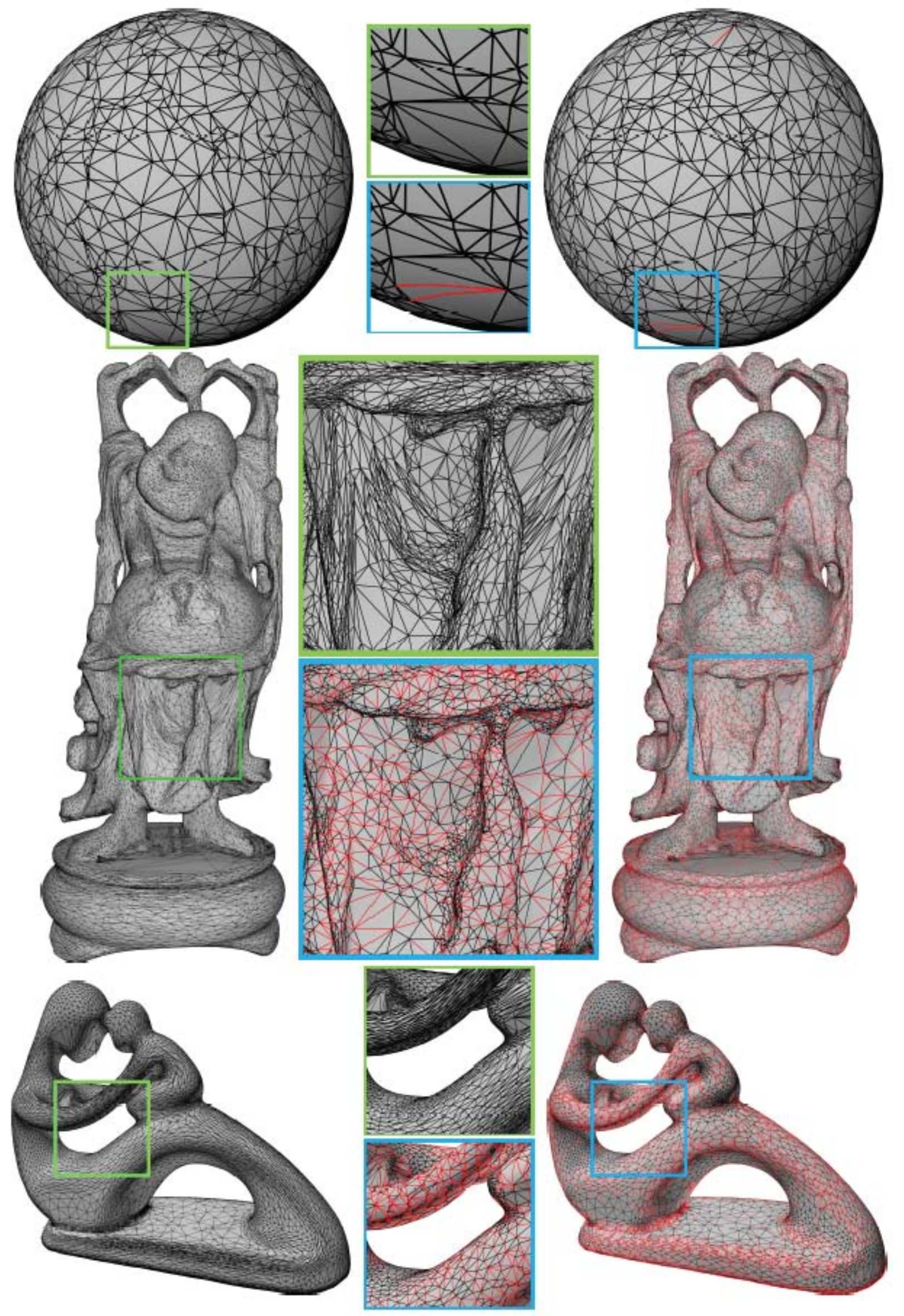}
  \vspace{-0.1in}
  \caption{Results. The original mesh edges and the Delaunay edges are colored in black and red, respectively.
  The figures are of high resolution, allowing close-up examination.}
  \label{fig:results}
  \end{figure}

\section{Experimental Results}
\label{sec:results}

  This section reports the experimental results and compare the performance of our method and the edge-flipping algorithm.

  We implement our algorithm in C++ and test it on 10 synthetic and real-world models with diverse geometric and topological features.
  As Table~\ref{table:runningtime} shows, most meshes are far from their Delaunay triangulations.
  For example, the Fertility model has 46.3\% non-Delaunay edges (see Figure~\ref{fig:results}),
  which takes the edge-flipping algorithm many iterations to fix them.
  Our algorithm, in contrast, computes the IDT in a non-iterative manner and its performance is not sensitive to the number of non-Delaunay edges.
  We observe that our method consistently outperforms the edge-flipping algorithm in terms of execution time.

  We also investigate the relation between mesh quality and performance.
  We measure the quality of a triangle $t$ by its {\it anisotropy} $\sigma=\frac{ph}{2\sqrt{3}S}$,
  where $p$ is the half-perimeter, $h$ is the length of its longest edge and $S$ is the triangle area.
  It is easy to verify that $\sigma(t)\geq 1$ and the equality holds when $t$ is equilateral.
  We measure the triangulation quality by the mean $\sigma_{mean}$, maximum $\sigma_{max}$ and standard deviation $\sigma_{std}$ of $\sigma$ for all triangles of $M$.
  Usually, a larger $\sigma$ means the higher degree of anisotropy and the further away the mesh is from its IDT,
  thus, the more edges that are flipped by the edge-flipping algorithm, and the higher the speedup our algorithm provides.

\begin{figure}[th]
\centering
\includegraphics[width=0.98\textwidth, bb=11 562 583 710, clip=true]{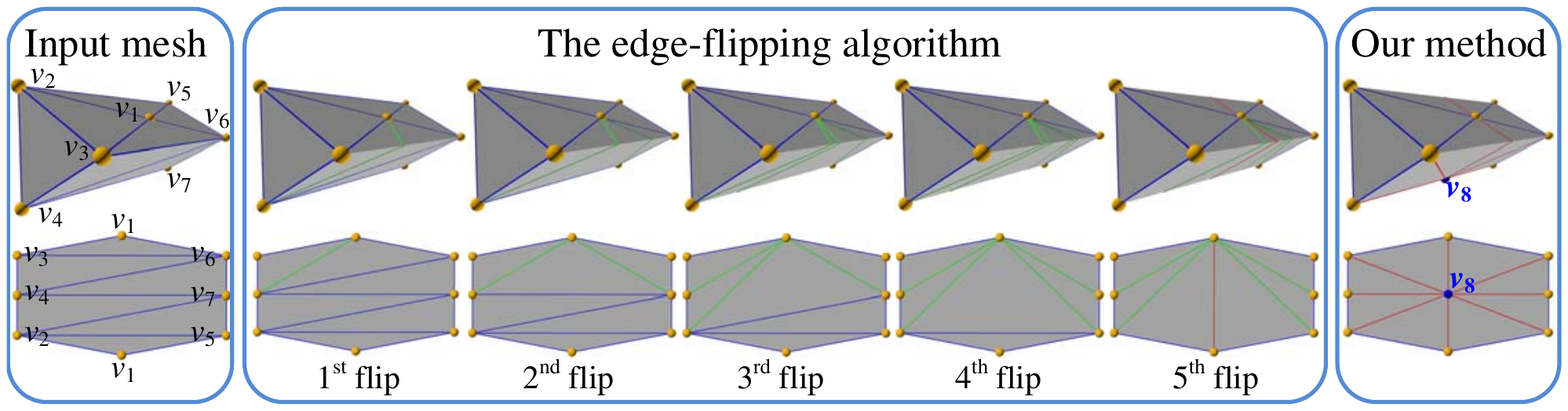}
\vspace{-0.1in}
\caption{Comparison with the edge-flipping algorithm. Consider a prism with 7 vertices. Among the four edges facing $v_1$, $v_2v_3$ and $v_5v_6$ are local Delaunay, whereas the other two edges $v_2v_5$ and $v_3v_6$ violate the local Delaunay condition. Thus, the edge-flipping algorithm iteratively flips the non-Delaunay edges until the local Delaunay property holds everywhere. However, the resulting Delaunay triangulation is not regular, due to the self-loop (in red) connecting $v_1$ and itself. Our method adds an auxiliary site $v_8$ which is the middle point of the self-loop and refines the Voronoi diagram locally. As a result, the dual Delaunay triangulation is regular. The bottom row shows the 2D flattening of the region in which the Delaunay property does not hold.}
\label{fig:edge-flip-horizontal2}
\end{figure}

\begin{figure}[th]
  \centering
  \includegraphics[width=0.6\textwidth, bb=52 422 523 696, clip=true]{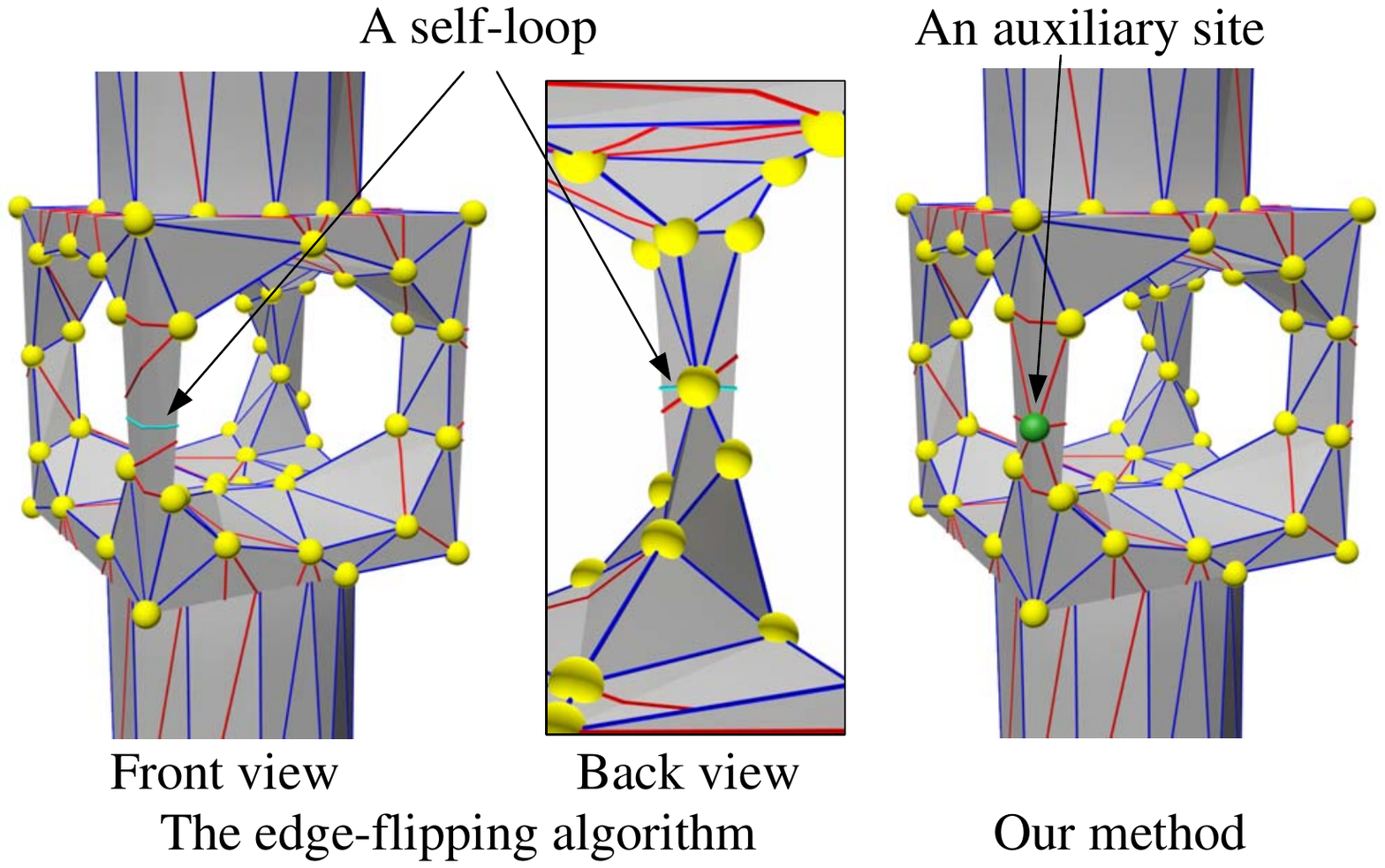}
  \vspace{-0.1in}
  \caption{The edge-flipping algorithm produces a self-loop on the CSG model (see the left and middle images).
  Our method solves the problem by adding an auxiliary site (see the green dot in the right image).
  Note that the original mesh edges (in blue) are line segments, whereas the IDT edges (in red) are geodesic paths.}
  \label{fig:real-comparison}
  \end{figure}

  As mentioned above, the edge-flipping algorithm may produce non-regular IDT.
  In contrast, our method guarantees the regular IDT by adding auxiliary sites at the poorly-sampled region.
  See Figures~\ref{fig:edge-flip-horizontal2} and \ref{fig:real-comparison}.
  Theoretically, our algorithm adds $O(n)$ auxiliary sites to ensure the existence of dual triangulation.
  In practice, we observe that only a very small number of auxiliary sites are required for real-world models.
  This is not a surprise, since the auxiliary sites are only added on the sparsely-sampled regions which are not homeomorphic to a disk
  (e.g., the cylinder-like geometry in Figures~\ref{fig:pipeline} and~\ref{fig:real-comparison}).
  As Table~\ref{table:runningtime} shows, although most test models are far from their Delaunay triangulations, they have a fairly good sampling density.
  Consequently, the resultant $GVD(V)$ automatically satisfies the closed ball density, and our algorithm does not add any new site at all.
  For these models, both our method and the edge-flipping algorithm produce exactly the same results.

\begin{figure}
  \centering
  \includegraphics[width=3.0in]{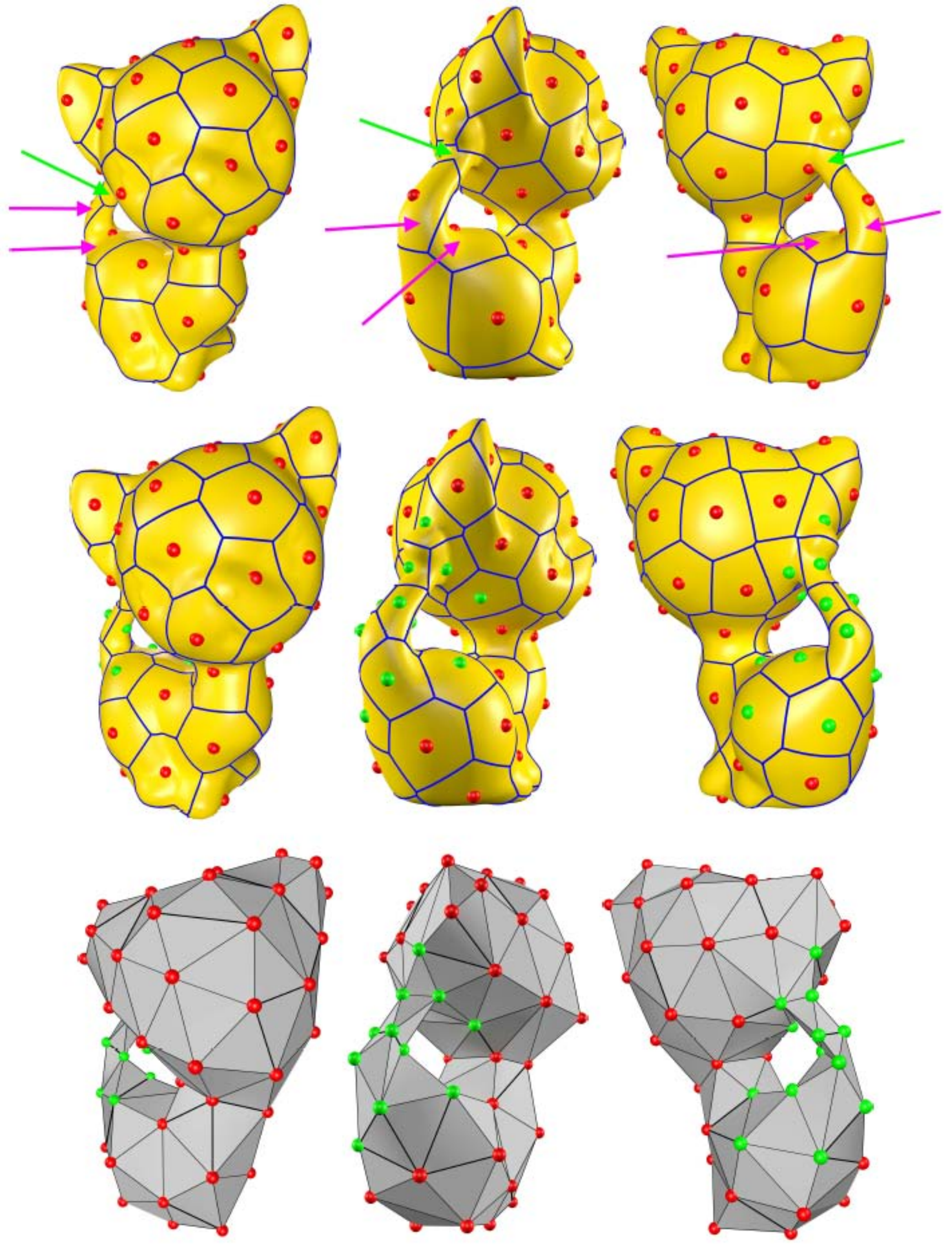}
  \vspace{-0.1in}
  \caption{Centroidal Voronoi tessellation.
  Row 1: The existing CVT algorithms (e.g.~\cite{Wang2015}) produce non-regular Voronoi cells when the generating sites are sparse.
  The green arrow indicates a multiply-connected Voronoi cell and the pink arrows show Voronoi cells sharing 2 common edges.
  Row 2: As a post-process, our method fixes the issues by adding auxiliary sites to the poorly-sampled regions and locally optimizing the site locations.
  Row 3: The dual graph of the updated CVT is a valid triangulation.
  The original sites and new sites are colored in red and green, respectively. }\label{fig:CVT}
  \end{figure}

  Our method can be easily adapted to centroidal Voronoi tessellation (CVT),
  which is a special type of Voronoi diagram such that the site of each Voronoi cell is also its center of mass.
  The existing work of CVT focuses on the convergence~\cite{DBLP:journals/siamnum/DuEJ06}, isotropic meshing~\cite{Alliez2005},
  energy functional smoothness~\cite{Liu2009CVT}, and intrinsic computation~\cite{Wang2015}.
  However, none of them addresses the issue of regularity.
  Indeed, all of them simply take it for granted that the site number is sufficiently large so that the CVT is regular.
  Unfortunately, when the generating sites are \textit{sparse}, non-regular Voronoi cells do exist.
  Figure~\ref{fig:CVT} shows an example of the genus-1 Kitten model with only 50 sites.
  Due to the low sampling density, the Voronoi cells around the tail do not satisfy the closed ball property.
  As a post-process to the existing techniques, our method automatically identifies the non-regular cells,
  adds auxiliary sites and then locally optimizes the site's location using the Lloyd method~\cite{DBLP:journals/siamnum/DuEJ06}.
  The updated CVT is regular, leading to a valid dual triangulation.

\section{Discrete Laplace-Beltrami Operators on Intrinsic Delaunay Triangulation}
\label{sec:LBO}

  The Laplace-Beltrami operator (LBO) is an important operator on Riemannian manifolds and it has many desired properties.
  For example, applying the LBO to the coordinate functions gives the mean curvature;
  the eigenfunctions of the Laplacian form a natural basis for square integrable functions on the manifold analogous to Fourier harmonics for functions on a circle.

\subsection{Convergence \& Accuracy}
\label{subsec:convergence}

  There is considerable amount of work of defining LBO on discrete domains~\cite{Pinkall1993,Dziuk1988,Desbrun1999,Meyer2003,Xu2004,Bobenko2007,DBLP:conf/sgp/WardetzkyMKG07,Belkin2008,pcdlp2009,DBLP:journals/tog/AlexaW11}.
  Most of these methods are variants of the \textit{cotangent scheme}~\cite{Pinkall1993}, which is a form of the finite element method applied to the Laplace operator on a surface:
  \begin{displaymath}
  \triangle f(v_i)=\sum_{\{v_i,v_j\}\in E}w_{ij}\left (f(v_i)-f(v_j)\right),
  \end{displaymath}
  where $f:V\rightarrow\mathbb{R}$ is a piecewise linear function defined on $M$.
  The weight for edge $e_{ij}=\{v_i,v_j\}$ is
  \begin{equation}
  \label{eqn:cotan}
  w_{ij}=\left\{
  \begin{array}{ll}
  \frac{1}{2}(\cot\alpha_{ij}+\cot\alpha_{ji}) & \mbox{if $e_{ij}$ is internal}\\
  \frac{1}{2}\cot\alpha_{ij} & \mbox{otherwise}
  \end{array}
  \right.
  \end{equation}
  where $\alpha_{ij}$ and $\alpha_{ji}$ are the two angles facing edge $e_{ij}$.
  The function $f$ is discrete harmonic, if $\bigtriangleup f(v_i)=0$ at all interior vertices.
  In spite of its extreme popularity in digital geometry processing, the cotangent formula has two serious drawbacks:
  \begin{itemize}
  \item~ The edge weight $w_{ij}$ in Equation (\ref{eqn:cotan}) could be negative, meaning that $f_i$ is not a \textit{convex} combination of its neighbors.
  In parameterization, this non-convex combination issue produces flipped triangles in the parametric domain.
  \item~ The formula is not intrinsic, that is, two surfaces that are isometric but with different triangulations may have different discrete LBOs.
  \end{itemize}
  Bobenko and Springborn \cite{Bobenko2007} proved that the cotangent weight is non-negative if the underlying triangulation is Delaunay.
  Moreover, since the intrinsic Delaunay tessellation is unique, the discrete LBO defined on $IDT(M)$ is also intrinsic and independent of the triangulation of $M$.

  In this subsection, we thoroughly evaluate the accuracy and convergence of commonly used discrete LBOs on the original mesh $M$ and the intrinsic Delaunay triangulation $IDT(M)$.

  \noindent (1) Classic cotangent LBO \cite{Pinkall1993}:
  \begin{displaymath}
  \triangle^{(1)}f(v_i)=\sum_{\{v_i,v_j\}\in E}w_{ij}(f(v_i)-f(v_j))
  \end{displaymath}
  where $w_{ij}$ is defined in Equation~(\ref{eqn:cotan}).

  \noindent (2) 1-ring-area-weighted cotangent LBO \cite{Desbrun1999}:
  \begin{displaymath}
  \triangle^{(2)}f(v_i)=\frac{1}{2A}\triangle^{(1)}f(v_i)
  \end{displaymath}
  where $A$ is the sum of areas of $v_i$'s 1-ring triangles.

  \noindent (3) Voronoi-area-weighted cotangent LBO \cite{Meyer2003}:
  \begin{displaymath}
  \triangle^{(3)}f(v_i)=\frac{1}{A_{voronoi}}\triangle^{(1)}f(v_i)
  \end{displaymath}
  where $A_{voronoi}$ is the Voronoi area at vertex $v_i$~\cite{DBLP:journals/cgf/GrinspunGRZ06}.

  \noindent (4) Mesh LBO \cite{Belkin2008}:
  \begin{displaymath}
  \triangle^{(4)}f(v_i)=\frac{1}{4\pi h^2}\sum_{f\in F}\frac{A(f)}{3}\sum_{v_j\in f}e^{-\frac{d^2(v_j,v_i)}{4h}}(f(v_i)-f(v_j)),
  \end{displaymath}
  where $A(f)$ is the area of triangle $f$. The parameter $h$ is a positive quantity, which intuitively corresponds to the size of the neighborhood considered at $v_i$.

  We test the above discrete LBOs on $\mathbb{R}^2$ and $\mathbb{S}^2$ where the analytical LBO is available:
  \begin{displaymath}
  \Delta_{\mathbb{R}^2}f(x,y)=\frac{\partial^2f}{\partial x^2}+\frac{\partial^2f}{\partial y^2}
  \end{displaymath}
  and
  \begin{displaymath}
  \Delta_{\mathbb{S}^2}f(\theta,\phi)=\frac{1}{\sin\theta}\frac{\partial}{\partial\theta}\left(\sin\theta\frac{\partial f}{\partial\theta}\right)+\frac{1}{\sin^2\theta}\frac{\partial^2 f}{\partial\phi^2},
  \end{displaymath}
  where $\theta$ and $\phi$ are the spherical coordinates.


  Following~\cite{Belkin2008}, we consider three functions on $\mathbb{R}^2$, $f_1(x,y)=x^2$, $f_2(x,y)=e^x$ and $f_3(x,y)=e^{x+y}$,
  and three functions on $\mathbb{S}^2$, $f_4(x,y,z)=x^2$, $f_5(x,y,z)=e^x$ and $f_6(x,y,z)=e^{x+y}$.
  We generate a sequence of triangle meshes with increasing resolution for each function $f(x,y)$.
  To obtain the planar meshes, we apply the greedy triangulation~\cite{greedy} to random samples in $[-0.5,0.5]\times[-0.5,0.5]$.
  We adopt the marching cube algorithm~\cite{Lorensen1987} to construct the spherical meshes,
  where the resolution is related to the user-specified cube size, i.e., the smaller cube size, the higher mesh resolution we obtain.
  As Table \ref{table:plane} shows, the IDTs induced LBOs produce smaller normalized $L_2$ error than those of the original meshes.
  The classic cotangent formula $\Delta^{(1)}$ does not converge to the analytical LBO at all,
  whereas the other three discrete LBOs converge.
  Moreover, when the mesh has a sufficiently high resolution, $\Delta^{(4)}$ has the least $L_2$ error.

  \begin{table}
  \begin{center}
  \begin{tabular}{|c|c|cc|cc|cc|cc|}
  \hline
  \multicolumn{10}{|c|}{Planar domain (unit square $[-0.5,0.5]\times[-0.5,0.5]$)}\\
  \hline
  \multicolumn{1}{|c|}{\multirow {3}{*}{Function}} & \multicolumn{1}{c|}{\multirow{3}{*}{$|E|$}} & \multicolumn{8}{c|}{Normalized $L_2$ error} \\
  \cline{3-10}
   & & \multicolumn{2}{|c|}{$\Delta^{(1)}$} & \multicolumn{2}{c|}{$\Delta^{(2)}$} & \multicolumn{2}{c|}{$\Delta^{(3)}$} & \multicolumn{2}{c|}{$\Delta^{(4)}$} \\
   \cline{3-10}
   &  & $M$ & IDT & $M$ & IDT & $M$ & IDT & $M$ & IDT\\
  \hline
  \multicolumn{1}{|c|}{\multirow {4}{*}{$f_1(x,y)$}} & 261 & 0.731 & 0.666 & 0.655 & 0.656 & 0.905 & \textcolor{red}{0.321} & 2.801 & 2.399\\
   & 1,121 & 0.803 & 0.844 & 0.604 & 0.595 & 0.391 & \textcolor{red}{0.183} & 0.587 & 0.513\\
   & 4,641 & 0.960 & 0.942 & 0.548 & 0.520 & 0.362 & \textcolor{red}{0.153} & 0.186 & 0.171\\
   & 18,881 & 0.961 & 0.960 & 0.469 & 0.417 & 0.255 & 0.119 & 0.041 & \textcolor{red}{0.040}\\
  \hline
  \multicolumn{1}{|c|}{\multirow {4}{*}{$f_2(x,y)$}} & 261 & 0.839 & 0.768 & 0.641 & 0.575 & 0.745 & \textcolor{red}{0.379} & 6.626 & 6.353\\
   & 1,121 & 0.956 & 0.879 & 0.640 & 0.571 & 0.452 & \textcolor{red}{0.160} & 1.248 & 1.058\\
   & 4,641 & 0.986 & 0.883 & 0.639 & 0.569 & 0.420 & \textcolor{red}{0.149} & 0.386 & 0.334\\
   & 18,881 & 0.996 & 0.874 & 0.632 & 0.543 & 0.334 & 0.135 & 0.082 & \textcolor{red}{0.074}\\
  \hline
  \multicolumn{1}{|c|}{\multirow {4}{*}{$f_3(x,y)$}} & 261 & 0.843 & 0.741 & 0.675 & 0.623 & 0.425 & \textcolor{red}{0.230} & 2.838 & 2.786\\
   & 1,121 & 0.951 & 0.849 & 0.647 & 0.621 & 0.373 & \textcolor{red}{0.161} & 1.305 & 1.225\\
   & 4,641 & 0.988 & 0.884 & 0.644 & 0.613 & 0.257 & \textcolor{red}{0.131} & 0.332 & 0.313\\
   & 18,881 & 0.997 & 0.924 & 0.637 & 0.578 & 0.200 & 0.105 & 0.076 & \textcolor{red}{0.067}\\
  \hline
  \hline
  \multicolumn{10}{|c|}{Spherical domain (unit sphere)}\\
  \hline
  \multicolumn{1}{|c|}{\multirow {3}{*}{Function}} & \multicolumn{1}{c|}{\multirow{3}{*}{$|E|$}}  & \multicolumn{8}{c|}{Normalized $L_2$ error} \\
  \cline{3-10}
   & & \multicolumn{2}{|c|}{$\Delta^{(1)}$} & \multicolumn{2}{c|}{$\Delta^{(2)}$} & \multicolumn{2}{c|}{$\Delta^{(3)}$} & \multicolumn{2}{c|}{$\Delta^{(4)}$} \\
   \cline{3-10}
   & & $M$ & IDT & $M$ & IDT & $M$ & IDT & $M$ & IDT\\
  \hline
  \multicolumn{1}{|c|}{\multirow {4}{*}{$f_4(x,y,z)$}} & 804 & 0.892 & 0.892 & 0.654 & 0.654 & 0.831 & \textcolor{red}{0.300} & 2.984 & 3.053\\
   & 3,468 & 0.972 & 0.972 & 0.642 & 0.642 & 0.432 & \textcolor{red}{0.190} & 0.561 & 0.545\\
   & 14,268 & 0.992 & 0.992 & 0.642 & 0.642 & 0.434 & 0.177 & 0.154 & \textcolor{red}{0.148}\\
   & 57,684 & 0.998 & 0.998 & 0.641 & 0.641 & 0.368 & 0.171 & \textcolor{red}{0.027} & \textcolor{red}{0.027}\\
  \hline
  \multicolumn{1}{|c|}{\multirow {4}{*}{$f_5(x,y,z)$}} & 804 & 0.671 & 0.697 & 0.648 & 0.663 & 0.817 & \textcolor{red}{0.394} & 6.930 & 6.804\\
   & 3,468 & 0.881 & 0.791 & 0.604 & 0.600 & 0.408 & \textcolor{red}{0.153} & 1.222 & 1.140\\
   & 14,268 & 0.893 & 0.877 & 0.543 & 0.536 & 0.341 & \textcolor{red}{0.119} & 0.319 & 0.330\\
   & 57,684 & 0.932 & 0.892 & 0.442 & 0.445 & 0.248 & 0.093 & \textcolor{red}{0.054} & \textcolor{red}{0.054}\\
  \hline
  \multicolumn{1}{|c|}{\multirow {4}{*}{$f_6(x,y,z)$}} & 804 & 0.652 & 0.701 & 0.722 & 0.699 & 0.396 & \textcolor{red}{0.211} & 3.508 & 2.858\\
   & 3,468 & 0.813 & 0.829 & 0.604 & 0.603 & 0.393 & \textcolor{red}{0.171} & 1.275 & 1.182\\
   & 14,268 & 0.859 & 0.878 & 0.540 & 0.555 & 0.316 & \textcolor{red}{0.156} & 0.280 & 0.268\\
   & 57,684 & 0.881 & 0.891 & 0.431 & 0.464 & 0.301 & 0.145 & \textcolor{red}{0.053} & \textcolor{red}{0.053}\\
  \hline
  \end{tabular}
  \end{center}
  \caption{Evaluation of various discrete LBOs on planar and spherical domains, which are triangulated in increasing resolutions. In each resolution, the domain with the least normalized $L_2$ error is shown in red. $|E|$ is the number of edges in the meshes.}
  \label{table:plane}
  \end{table}

\subsection{Mean Curvature Computation}
\label{subsec:mean_curvature}

  We evaluate the performance of various discrete LBOs on mean curvature computation.
  Applying the LBO to the coordinate functions of a point $p\in S$, we obtain the mean curvature vector, i.e.,
  \begin{displaymath}
  \Delta p=2H(p)\mathbf{n}(p)
  \end{displaymath}
  where $H(p)$ and $\mathbf{n}(p)$ are the mean curvature and unit outward normal at $p$.
  Following \cite{Xu2004}, we consider smooth surfaces given by the following non-linear functions:
  \begin{displaymath}
  F_1(x,y)=\sqrt{4-(x-0.5)^2-(y-0.5)^2},
  \end{displaymath}
  \begin{displaymath}
  F_2(x,y)=\tanh(9y-9x),
  \end{displaymath}
  \begin{displaymath}
  F_3(x,y)=\frac{1.25+\cos(5.4y)}{6+6(3x-1)^2},
  \end{displaymath}
  \begin{displaymath}
  F_4(x,y)=e^{-\frac{81}{16}((x-0.5)^2+(y-0.5)^2)}
  \end{displaymath}

  Similar to the convergence test, we generate a sequence of triangle meshes with increasing resolutions for each smooth surface $F(x,y)$.
  Then we evaluate the accuracy of the mean curvature computed by using the conventional LBOs and the IDT induced LBOs.
  As Table~\ref{table:mean_curvature} shows, the IDT induced LBOs produce more accurate results than the LBOs defined on the original meshes.

  \begin{table}
  \begin{center}
  \begin{tabular}{|c|c|cc|cc|cc|cc|}
  \hline
  \multicolumn{1}{|c|}{\multirow {3}{*}{Function}} & \multicolumn{1}{c|}{\multirow{3}{*}{$|E|$}}  & \multicolumn{8}{c|}{Normalized $L_2$ error} \\
  \cline{3-10}
   &  & \multicolumn{2}{|c|}{$\Delta^{(1)}$} & \multicolumn{2}{c|}{$\Delta^{(2)}$} & \multicolumn{2}{c|}{$\Delta^{(3)}$} & \multicolumn{2}{c|}{$\Delta^{(4)}$} \\
  \cline{3-10}
   & & $M$ & IDT & $M$ & IDT & $M$ & IDT & $M$ & IDT\\
  \hline
  \multicolumn{1}{|c|}{\multirow {5}{*}{$F_1(x,y)$}} & 539 & 0.861 & 0.805 & 0.580 & 0.491 & 0.081 & \textcolor{red}{0.039} & 2.378 & 2.023\\
   & 2,311 & 0.868 & 0.819 & 0.580 & 0.443 & 0.078 & \textcolor{red}{0.033} & 2.362 & 1.765\\
   & 9,423 & 0.869 & 0.783 & 0.580 & 0.354 & 0.078 & \textcolor{red}{0.029} & 2.324 & 1.455\\
   & 38,177 & 0.870 & 0.820 & 0.580 & 0.319 & 0.078 & \textcolor{red}{0.024} & 2.235 & 1.317\\
   & 153,095 & 0.870 & 0.818 & 0.580 & 0.253 & 0.078 & \textcolor{red}{0.019} & 2.004 & 0.953\\
  \hline
  \multicolumn{1}{|c|}{\multirow {5}{*}{$F_2(x,y)$}} & 334 & 0.691 & 0.655 & 0.522 & 0.469 & 0.253 & \textcolor{red}{0.186} & 0.661 & 0.535\\
   & 1,666 & 0.698 & 0.643 & 0.481 & 0.351 & 0.080 & \textcolor{red}{0.048} & 0.654 & 0.473\\
   & 7,372 & 0.699 & 0.657 & 0.470 & 0.310 & 0.045 & \textcolor{red}{0.024} & 0.648 & 0.447\\
   & 30,655 & 0.700 & 0.652 & 0.468 & 0.255 & 0.039 & \textcolor{red}{0.019} & 0.635 & 0.370\\
   & 124,431 & 0.700 & 0.673 & 0.467 & 0.192 & 0.038 & \textcolor{red}{0.014} & 0.628 & 0.270\\
  \hline
  \multicolumn{1}{|c|}{\multirow {5}{*}{$F_3(x,y)$}} & 547 & 0.752 & 0.721 & 0.519 & 0.458 & 0.064 & \textcolor{red}{0.056} & 0.854 & 0.752\\
   & 2,677 & 0.758 & 0.691 & 0.510 & 0.391 & 0.049 & \textcolor{red}{0.028} & 0.848 & 0.606\\
   & 11,469 & 0.759 & 0.751 & 0.508 & 0.332 & 0.039 & \textcolor{red}{0.025} & 0.832 & 0.572\\
   & 47,437 & 0.760 & 0.718 & 0.507 & 0.291 & 0.041 & \textcolor{red}{0.020} & 0.793 & 0.447\\
   & 192,825 & 0.760 & 0.699 & 0.507 & 0.203 & 0.041 & \textcolor{red}{0.014} & 0.683 & 0.328\\
  \hline
  \multicolumn{1}{|c|}{\multirow {5}{*}{$F_4(x,y)$}} & 401 & 0.825 & 0.752 & 0.693 & 0.553 & 0.473 & \textcolor{red}{0.401} & 0.712 & 0.633\\
   & 1,687 & 0.829 & 0.780 & 0.684 & 0.509 & 0.462 & \textcolor{red}{0.333} & 0.700 & 0.501\\
   & 7,117 & 0.830 & 0.765 & 0.682 & 0.436 & 0.461 & \textcolor{red}{0.276} & 0.675 & 0.470\\
   & 29,265 & 0.830 & 0.800 & 0.682 & 0.362 & 0.460 & \textcolor{red}{0.252} & 0.626 & 0.338\\
   & 118,673 & 0.830 & 0.789 & 0.682 & 0.309 & 0.459 & \textcolor{red}{0.195} & 0.553 & 0.234\\
  \hline
  \end{tabular}
  \end{center}
  \caption{Mean curvature computation.
  Among the four discrete LBOs, the Voronoi-area-weighted LBO defined on the IDT (in red) produces the most accurate results.
  $|E|$ is the number of edges in the meshes.}
  \label{table:mean_curvature}
  \end{table}

\subsection{Applications}
\label{subsec:applications}

  As mentioned above, the IDT induced cotangent LBO is intrinsic to the geometry and non-negative for all edges.
  These features are highly desirable to many graphics applications, such as denoising~\cite{Desbrun1999},
  parameterization~\cite{Gu2003,DBLP:journals/cgf/KalbererNP07}, quadrangulation~\cite{DBLP:journals/tog/BommesZK09}, manifold harmonics~\cite{DBLP:journals/cgf/ValletL08},
  shape signature~\cite{DBLP:journals/cgf/SunOG09}, diffusion distance~\cite{diffusion} and biharmonic distance~\cite{DBLP:journals/tog/LipmanRF10},
  just name a few.
  In this subsection, we demonstrate IDT on harmonic mapping and manifold harmonics, two typical applications based on the discrete LBO.
  The conformal discrete Laplacian matrix $\mathbf{L}_c$~\cite{Pinkall1993} is
  \begin{displaymath}
  \mathbf{L}_{ij}=\left\{
  \begin{array}{cl}
  \sum_{k} w_{ik}, & {\rm if}~ i=j\\
  -w_{ij}, & {\rm if}~  \{v_i, v_j\}\in E \\
  0 & {\rm otherwise}
  \end{array} \right.
  \end{displaymath}
  where $w_{ij}$ is the cotangent weight for edge $\{v_i,v_j\}$ (see Equation~(\ref{eqn:cotan})).
  Let $\mathbf{A}$ be a diagonal matrix, where $\mathbf{A}_{ii}$ is the Voronoi area at vertex $v_i$.
  Then the discrete Laplacian matrix $\mathbf{L}$~\cite{Meyer2003} is given by
  \begin{displaymath}
  \mathbf{L}=\mathbf{A}^{-1}\mathbf{L}_c.
  \end{displaymath}

  The harmonic mapping is to solve a linear system of $\mathbf{L}$. Let $\phi:V\rightarrow\mathbb{R}$ be a scalar function defined on mesh vertices.
  Since $rank(\mathbf{L})=n-1$, we need to fix at least one vertex to get a unique solution.
  Without loss of generality, say $\phi(v_1)=0$. The function $\phi$ is harmonic if $\bigtriangleup \phi(v_i)=0$ for all vertices $v_i$, $i>1$.
  The discrete harmonic map $\phi$ is realized by solving the linear system $\widetilde{\mathbf{L}}\boldsymbol{\phi}=\mathbf{b}$,
  where $\boldsymbol{\phi}=(\ldots,\phi(v_i),\ldots)^T$, $\mathbf{b}=(1,0,0,\cdots)^{T}\in\mathbb{R}^{n\times 1}$,
  and matrix $\widetilde{\mathbf{L}}$ coincides with the discrete Laplacian matrix $\bf L$ except for the first row, which is $(1, 0, 0, \cdots)$.
  We compute the condition number of $\widetilde{\mathbf{L}}$, which is a good measure of the numerical stability of the linear system.
  As Table~\ref{table:runningtime} shows, the IDT induced matrix $\widetilde{\mathbf{L}}$ has smaller condition number than the matrix produced by the original mesh.

\begin{figure}[th]
  \centering
  \includegraphics[width=1.05in]{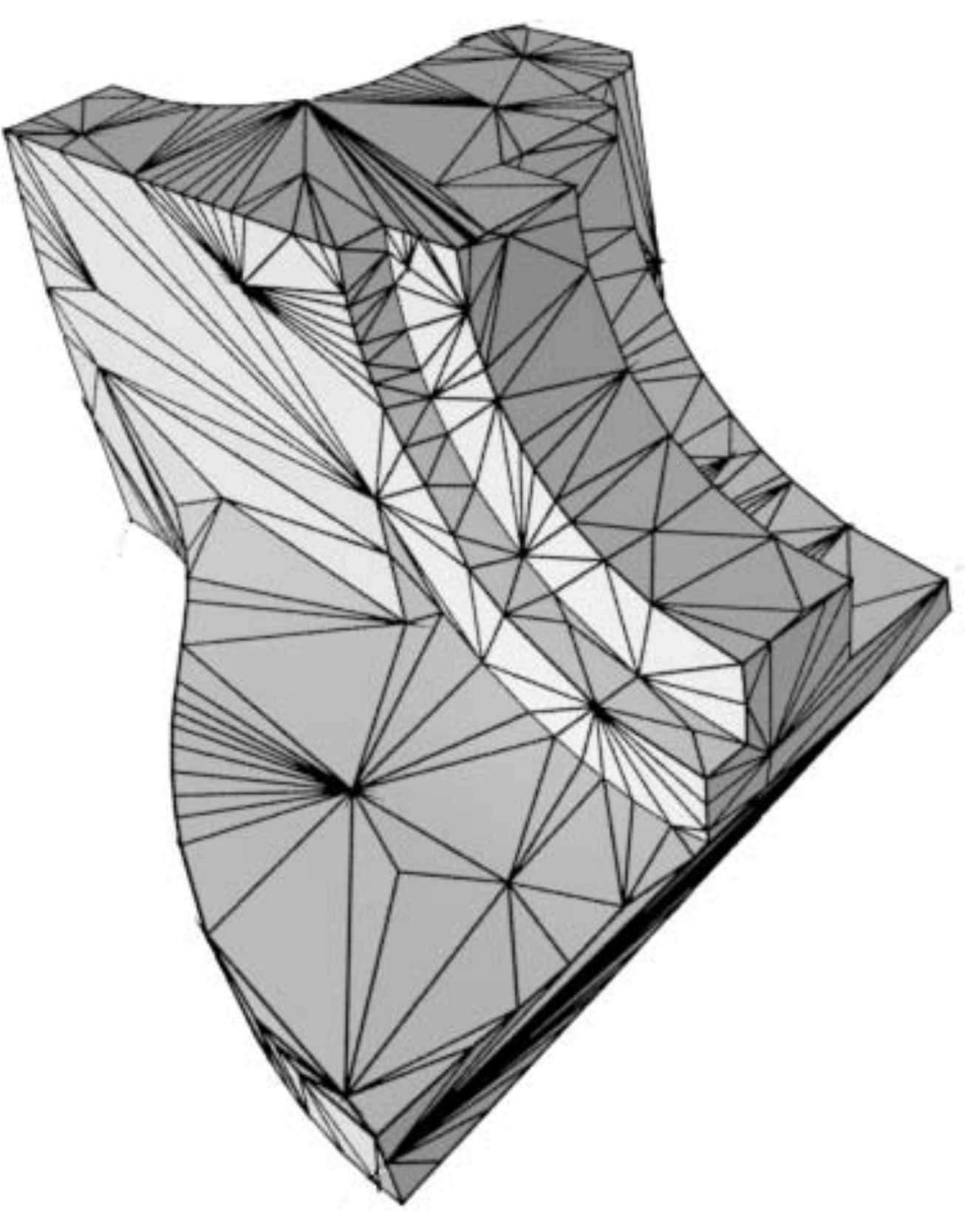}
  \includegraphics[width=1.05in]{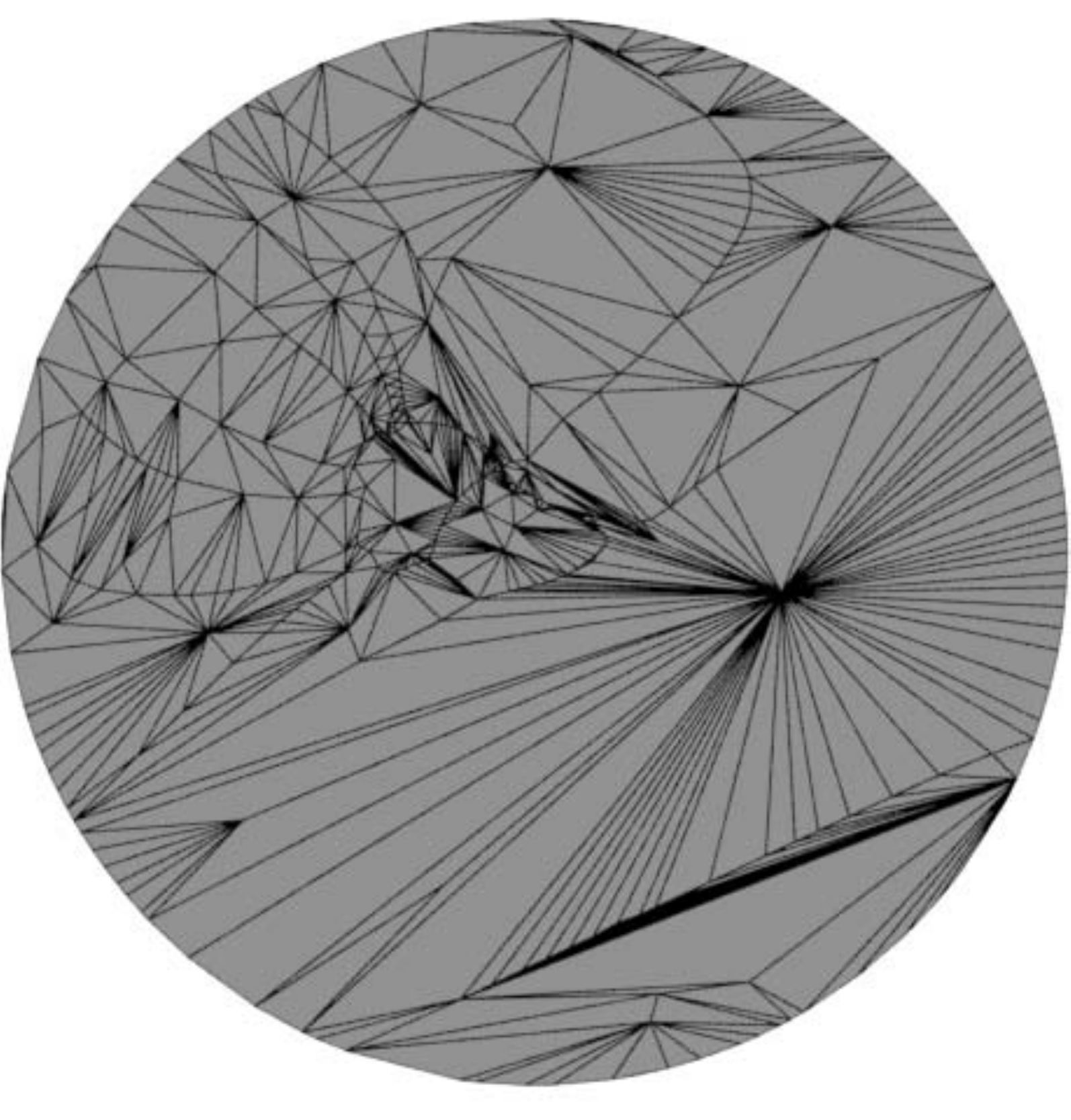}
  \includegraphics[width=1.05in]{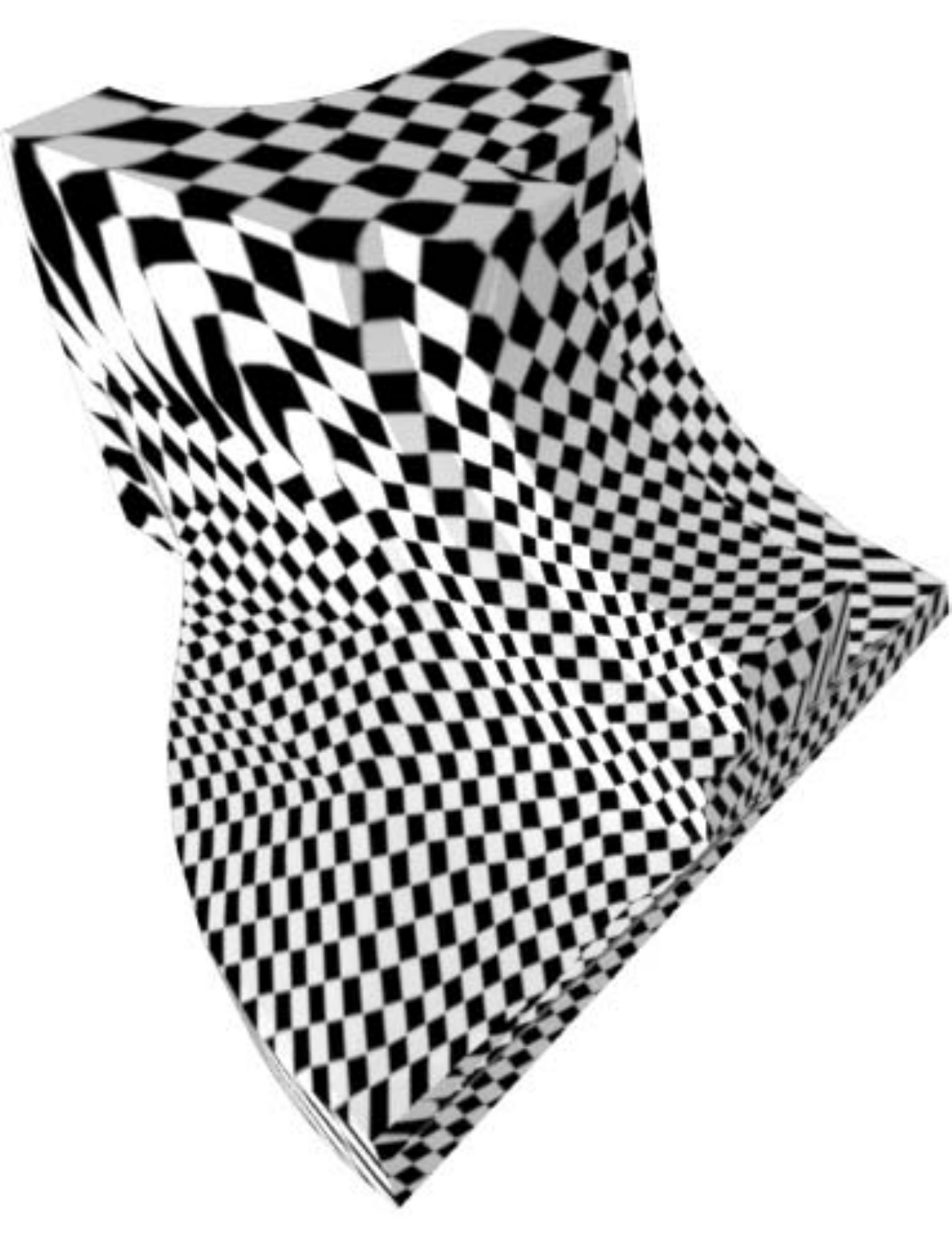}\\
  \includegraphics[width=1.05in]{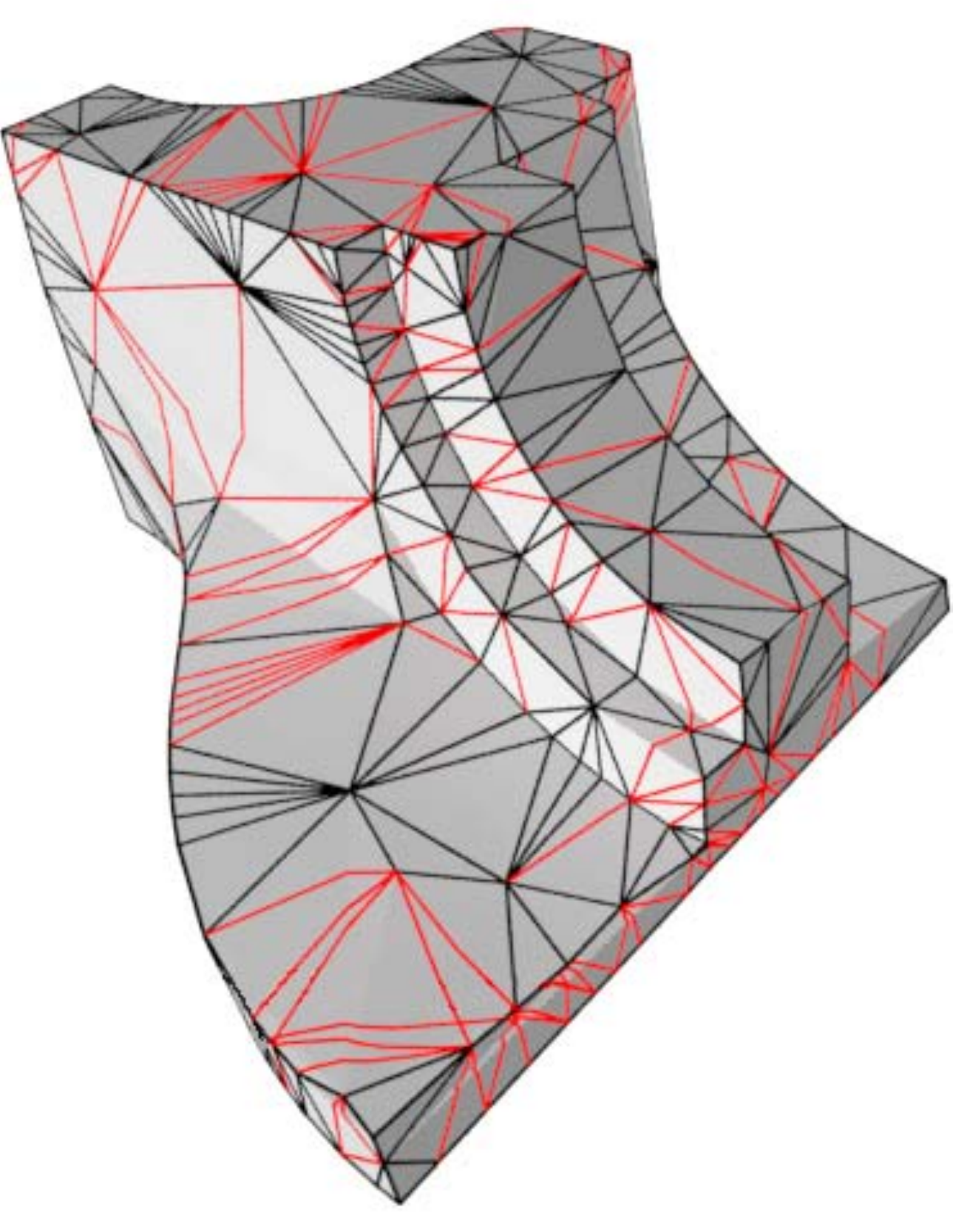}
  \includegraphics[width=1.05in]{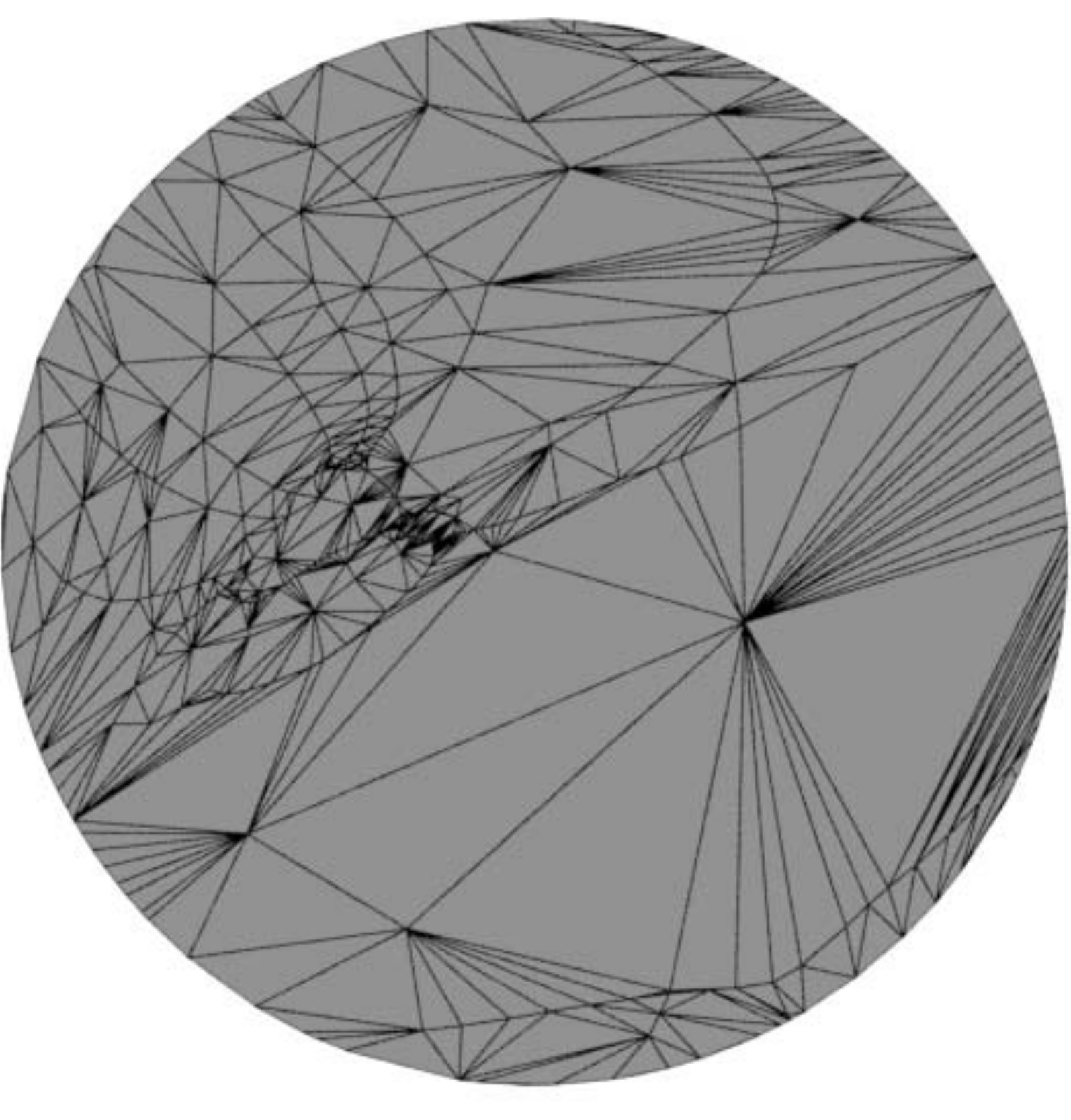}
  \includegraphics[width=1.05in]{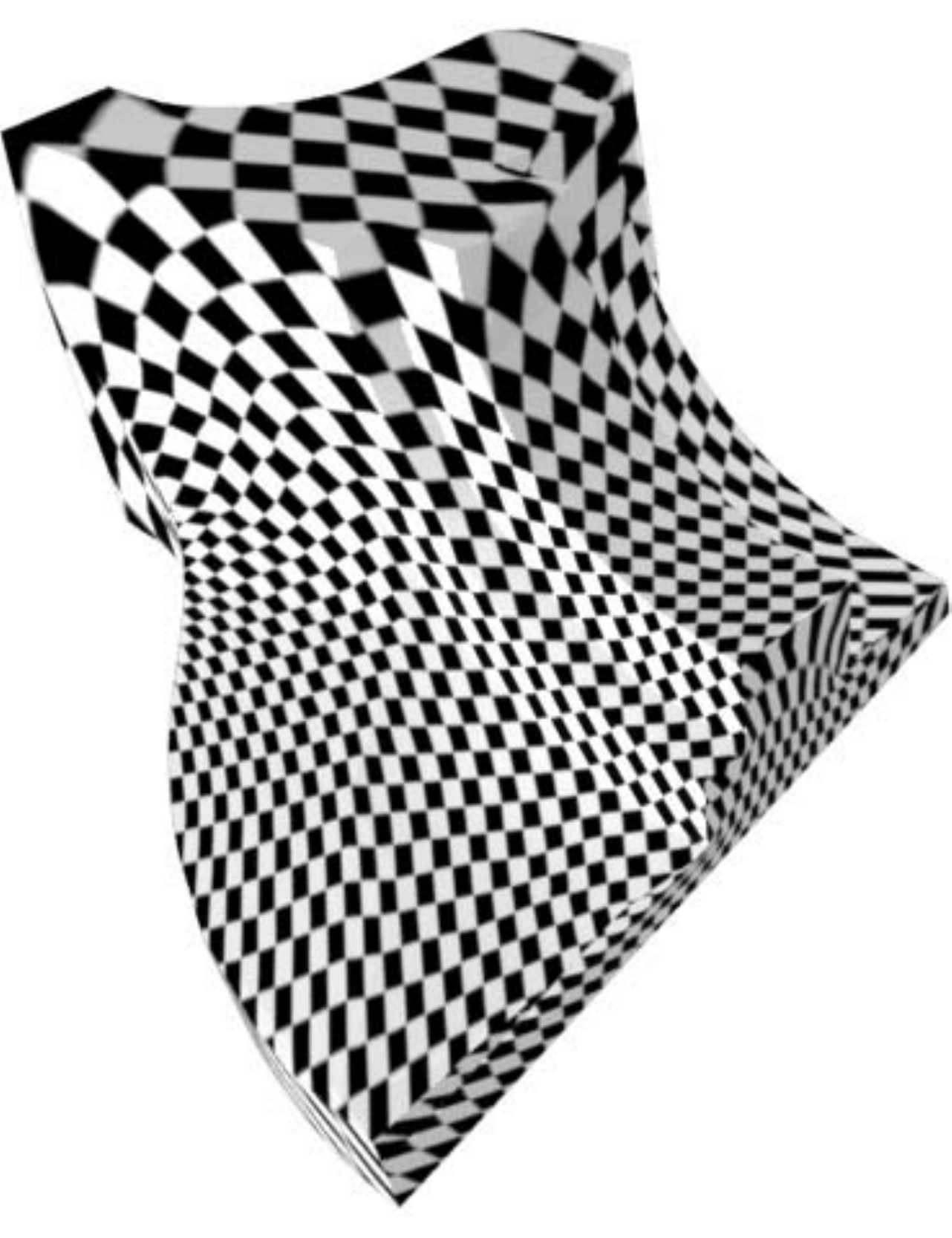}\\
  \makebox[0.7in]{(a) Triangulation} \makebox[1.5in]{(b) 2D domain}\makebox[1.1in]{(c) Texture mapping}
    \caption{We parameterize the Fandisk model to 2D unit disk using harmonic map.
  The original mesh (row 1) has many skinny triangles, so there are many negative weights in the cotangent LBO.
  Our IDT (row 2) guarantees non-negative weights for all internal edges and enforces the convex combination for all internal vertices.
  As a result, the IDT based paramterization is more stable and has less artifact than the original mesh.
  The parameterizations of $M$ and $IDT(M)$ have angle distortion 1.70 and 1.55, respectively.}
  \label{fig:fandisk}
  \end{figure}

  We use harmonic mapping to parameterize the Fandisk model, which is a genus-0 model with 1 boundary (i.e., a topological disk).
  The boundary vertices are mapped to unit circle $\mathbb{S}^1$ using arc-length parameterization (see Figure~\ref{fig:fandisk}(b)).
  We evaluate the parameterization quality by measuring the angle distortion
  \begin{displaymath}
  \epsilon=\sum_{f\in F}\frac{\cot\alpha a^2+\cot\beta b^2+\cot\gamma c^2}{4A(f)}A(f'),
  \end{displaymath} where $a$, $b$, $c$, $\alpha$, $\beta$, $\gamma$ are the side lengths and angles of triangle $f\in F$,
  $f'\in\mathbb{R}^2$ is the parameterized 2D triangle of $f$, and $A(\cdot)$ is the triangle area.
  To visualize $\phi$ on the IDT using texture mapping,
  we tessellate each non-planar geodesic triangle to planar polygons and linearly interpolate the texture coordinates for points
  at which the geodesic edges and the mesh edges meet.
  As Figure~\ref{fig:fandisk} shows, the IDT based harmonic mapping is numerically more stable than the original mesh.

\begin{figure}
  \centering
  \includegraphics[width=5.9in]{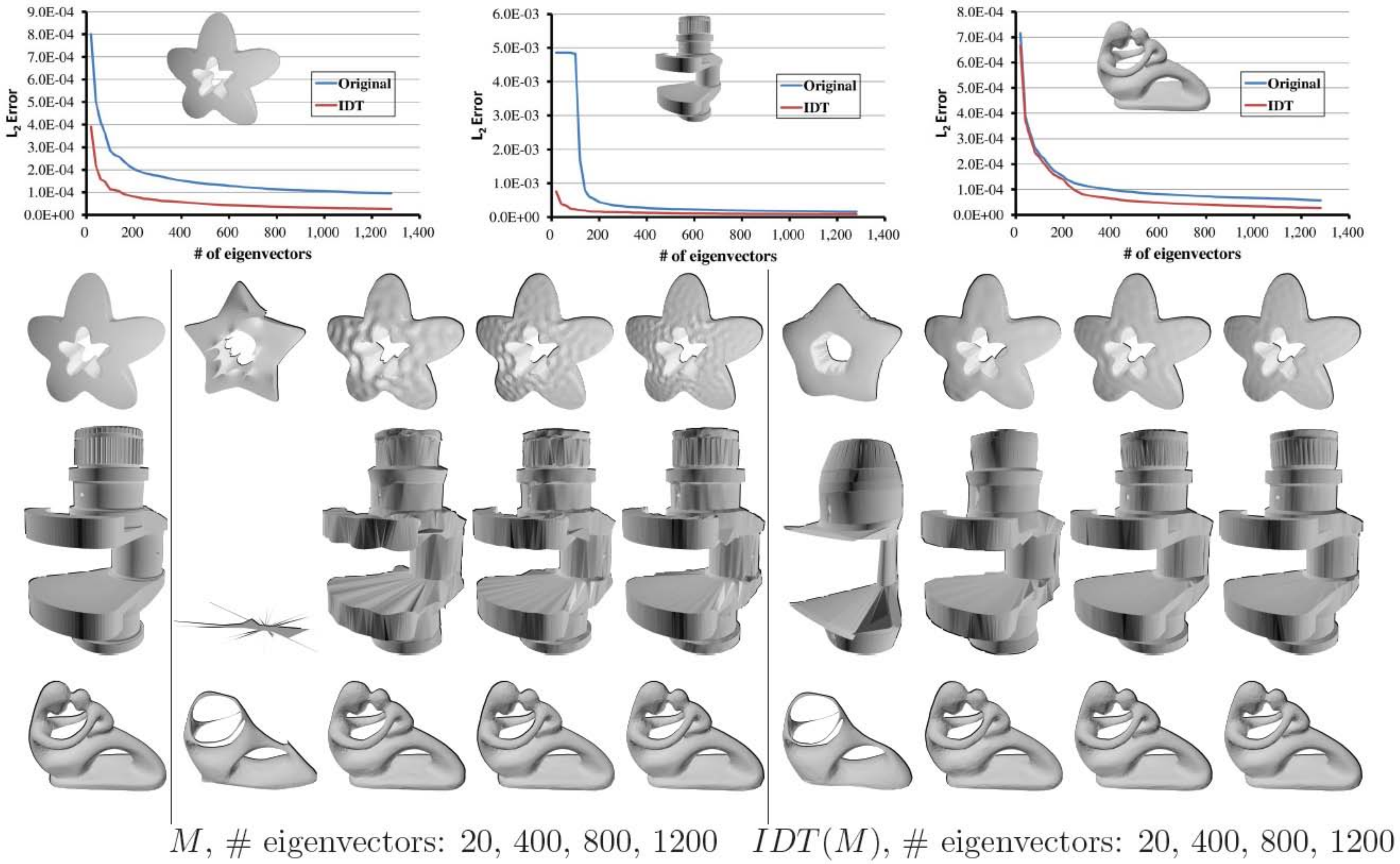}
  \vspace{-0.1in}
  \caption{Manifold harmonics. Row 1: The curve plots show that the IDT induced discrete Laplacian matrix produces more accurate results than the original mesh.
  Row 2: The visual results also confirm the IDTs have less artifacts than the original meshes, especially when only a small number of eigenvectors are used for reconstruction. }
  \label{fig:harmonics}
  \end{figure}

  Manifold harmonics~\cite{DBLP:journals/cgf/ValletL08} are a natural generalization of the Fourier transform to curved domains.
  Since the conformal discrete Laplacian matrix $\mathbf{L}_c$ is real symmetric, its eigenvalues are real and its eigenvectors are orthogonal,
  which define a function basis allowing for such a transform.
  We transform the geometry into frequency space by projecting the coordinate functions onto the eigenvectors corresponding to low frequency.
  Then we transform the object back into geometry space using the inverse transform.
  Figure~\ref{fig:harmonics} shows the geometry reconstructed from manifold harmonics with an increasing number of eigenvectors.
  We observe that the IDT based manifold harmonics have less artifacts than those based on the original meshes,
  especially when only a small number of eigenvectors are used for reconstruction.

\section{Conclusions}
\label{sec:conclusion}


\begin{thebibliography}{10}

\bibitem{Okabe2000}
A.~Okabe, B.~Boots, K.~Sugihara, and S.-N. Chiu, {\em Spatial Tessellations:
  Concept and Applications of {V}oronoi Diagrams}.
\newblock Wiley, 2000.

\bibitem{Edelsbrunner1997}
H.~Edelsbrunner and N.~R. Shah, ``Triangulating topological spaces,'' {\em
  International Journal of Computational Geometry and Applications}, vol.~7,
  no.~4, pp.~365--378, 1997.

\bibitem{Dyer2008}
R.~Dyer, H.~Zhang, and T.~M\"{o}ller, ``Surface sampling and the intrinsic
  {V}oronoi diagram,'' in {\em Proceedings of the Symposium on Geometry
  Processing}, pp.~1393--1402, 2008.

\bibitem{Boissonnat2013}
J.~Boissonnat, R.~Dyer, and A.~Ghosh, ``Constructing intrinsic delaunay
  triangulations of submanifolds,'' {\em CoRR}, vol.~abs/1303.6493, 2013.

\bibitem{rivin1994}
I.~Rivin, ``Euclidean structures on simplicial surfaces and hyperbolic
  volume,'' {\em Annals of Mathematics}, vol.~139, no.~3, pp.~553--580, 1994.

\bibitem{Indermitte2001}
C.~Indermitte, T.~Liebling, M.~Troyanov, and H.~Cl{\'e}men\c{c}on, ``{V}oronoi
  diagrams on piecewise flat surfaces and an application to biological
  growth,'' {\em Theoretical Computer Science}, vol.~263, July 2001.

\bibitem{Bobenko2007}
A.~I. Bobenko and B.~A. Springborn, ``A discrete laplace╟beltrami operator for
  simplicial surfaces,'' {\em Discrete \& Computational Geometry}, vol.~38,
  no.~4, pp.~740--756, 2007.

\bibitem{Fisher2006}
M.~Fisher, B.~Springborn, A.~I. Bobenko, and P.~Schr\"{o}der, ``An algorithm
  for the construction of intrinsic {D}elaunay triangulations with applications
  to digital geometry processing,'' {\em Computing}, vol.~82, no.~2-3,
  pp.~199--213, 2007.

\bibitem{Liu2011}
Y.-J. Liu, Z.~Chen, and K.~Tang, ``Construction of iso-contours, bisectors, and
  {V}oronoi diagrams on triangulated surfaces,'' {\em IEEE Transactions on
  Pattern Analysis and Machine Intelligence}, vol.~33, no.~8, pp.~1502--1517,
  2011.

\bibitem{Mitchell1987}
J.~S. Mitchell, D.~M. Mount, and C.~H. Papadimitriou, ``The discrete geodesic
  problem,'' {\em SIAM Journal on Computing}, vol.~16, no.~4, pp.~647--668,
  1987.

\bibitem{Ying13}
X.~Ying, X.~Wang, and Y.~He, ``Saddle vertex graph ({SVG}): A novel solution to
  the discrete geodesic problem,'' {\em ACM Transactions on Graphics},
  pp.~170:1--12, 2013.

\bibitem{Leibon2000}
G.~Leibon and D.~Letscher, ``{D}elaunay triangulations and {V}oronoi diagrams
  for {R}iemannian manifolds,'' in {\em Proceedings of the Sixteenth Annual
  Symposium on Computational Geometry}, pp.~341--349, ACM, 2000.

\bibitem{Xu2014}
C.~Xu, Y.-J. Liu, Q.~Sun, J.~Li, and Y.~He, ``Polyline-sourced geodesic
  {V}oronoi diagrams on triangle meshes,'' {\em Computer Graphics Forum},
  vol.~33, no.~7, pp.~161--170, 2014.

\bibitem{Edelsbrunner1990}
H.~Edelsbrunner and E.~P. M\"{u}cke, ``Simulation of simplicity: A technique to
  cope with degenerate cases in geometric algorithms,'' {\em ACM Trans.
  Graph.}, vol.~9, no.~1, pp.~66--104, 1990.

\bibitem{Mitchell87}
J.~Mitchell, D.~Mount, and C.~Papadimitriou, ``The discrete geodesic problem,''
  {\em SIAM Journal on Computing}, vol.~16, no.~4, pp.~647--668, 1987.

\bibitem{DBLP:journals/siamnum/DuEJ06}
Q.~Du, M.~Emelianenko, and L.~Ju, ``Convergence of the {L}loyd algorithm for
  computing centroidal {V}oronoi tessellations,'' {\em SIAM J. Numerical
  Analysis}, vol.~44, no.~1, pp.~102--119, 2006.

\bibitem{Alliez2005}
P.~Alliez, \'{E}ric Colin~de Verdi\`{e}re, O.~Devillers, and M.~Isenburg,
  ``Centroidal {V}oronoi diagrams for isotropic surface remeshing,'' {\em
  Graphical Models}, vol.~67, no.~3, pp.~204--231, 2005.

\bibitem{Liu2009CVT}
Y.~Liu, W.~Wang, B.~L{\'e}vy, F.~Sun, D.-M. Yan, L.~Lu, and C.~Yang, ``On
  centroidal {V}oronoi tessellation: Energy smoothness and fast computation,''
  {\em ACM Trans. Graph.}, vol.~28, no.~4, pp.~101:1--101:17, 2009.

\bibitem{Wang2015}
X.~Wang, X.~Ying, Y.-J. Liu, S.-Q. Xin, W.~Wang, X.~Gu, W.~Mueller-Wittig, and
  Y.~He, ``Intrinsic computation of centroidal {V}oronoi tessellation ({CVT})
  on meshes,'' {\em Computer-Aided Design}, vol.~58, pp.~51--61, 2015.

\bibitem{Pinkall1993}
U.~Pinkall and K.~Polthier, ``Computing discrete minimal surfaces and their
  conjugates,'' {\em Experimental Mathematics}, vol.~2, no.~1, pp.~15--36,
  1993.

\bibitem{Dziuk1988}
G.~Dziuk, ``Finite elements for the {B}eltrami operator on arbitrary
  surfaces,'' in {\em Partial Differential Equations and Calculus of
  Variations}, pp.~142--155, 1988.

\bibitem{Desbrun1999}
M.~Desbrun, M.~Meyer, P.~Schr\"{o}der, and A.~H. Barr, ``Implicit fairing of
  irregular meshes using diffusion and curvature flow,'' in {\em ACM SIGGRAPH},
  pp.~317--324, 1999.

\bibitem{Meyer2003}
M.~Meyer, M.~Desbrun, P.~Schr\"{o}der, and A.~H. Barr, ``Discrete
  differential-geometry operators for triangulated 2-manifolds,'' in {\em
  Visualization and Mathematics III}, pp.~35--57, 2003.

\bibitem{Xu2004}
G.~Xu, ``Convergence of discrete {L}aplace-{B}eltrami operators over
  surfaces,'' {\em Computers \& Mathematics with Applications}, vol.~48,
  no.~3-4, pp.~347--360, 2004.

\bibitem{DBLP:conf/sgp/WardetzkyMKG07}
M.~Wardetzky, S.~Mathur, F.~K{\"{a}}lberer, and E.~Grinspun, ``Discrete laplace
  operators: no free lunch,'' in {\em Proceedings of the Fifth Eurographics
  Symposium on Geometry Processing}, pp.~33--37, 2007.

\bibitem{Belkin2008}
M.~Belkin, J.~Sun, and Y.~Wang, ``Discrete {L}aplace operator on meshed
  surfaces,'' in {\em SCG '08}, pp.~278--287, 2008.

\bibitem{pcdlp2009}
M.~Belkin, J.~Sun, and Y.~Wang, ``Constructing laplace operator from point
  clouds in $\mathbb{R}^{d}$,'' in {\em SODA '09}, pp.~1031--1040, 2009.

\bibitem{DBLP:journals/tog/AlexaW11}
M.~Alexa and M.~Wardetzky, ``Discrete laplacians on general polygonal meshes,''
  {\em {ACM} Trans. Graph.}, vol.~30, no.~4, p.~102, 2011.

\bibitem{DBLP:journals/cgf/GrinspunGRZ06}
E.~Grinspun, Y.~I. Gingold, J.~Reisman, and D.~Zorin, ``Computing discrete
  shape operators on general meshes,'' {\em Comput. Graph. Forum}, vol.~25,
  no.~3, pp.~547--556, 2006.

\bibitem{greedy}
F.~P. Preparata and M.~I. Shamos, {\em Computational Geometry: An
  Introduction}.
\newblock Springer-Verlag New York Inc., 1988.

\bibitem{Lorensen1987}
W.~E. Lorensen and H.~E. Cline, ``Marching cubes: A high resolution 3{D}
  surface construction algorithm,'' in {\em Proceedings of SIGGRAPH '87},
  pp.~163--169, 1987.

\bibitem{Gu2003}
X.~Gu and S.-T. Yau, ``Global conformal surface parameterization,'' in {\em
  Proceedings of Symposium on Geometry Processing (SGP '03)}, pp.~127--137,
  2003.

\bibitem{DBLP:journals/cgf/KalbererNP07}
F.~K{\"{a}}lberer, M.~Nieser, and K.~Polthier, ``Quadcover - surface
  parameterization using branched coverings,'' {\em Comput. Graph. Forum},
  vol.~26, no.~3, pp.~375--384, 2007.

\bibitem{DBLP:journals/tog/BommesZK09}
D.~Bommes, H.~Zimmer, and L.~Kobbelt, ``Mixed-integer quadrangulation,'' {\em
  {ACM} Trans. Graph.}, vol.~28, no.~3, 2009.

\bibitem{DBLP:journals/cgf/ValletL08}
B.~Vallet and B.~L{\'{e}}vy, ``Spectral geometry processing with manifold
  harmonics,'' {\em Comput. Graph. Forum}, vol.~27, no.~2, pp.~251--260, 2008.

\bibitem{DBLP:journals/cgf/SunOG09}
J.~Sun, M.~Ovsjanikov, and L.~J. Guibas, ``A concise and provably informative
  multi-scale signature based on heat diffusion,'' {\em Comput. Graph. Forum},
  vol.~28, no.~5, pp.~1383--1392, 2009.

\bibitem{diffusion}
R.~Coifman and S.~Lafon, ``Diffusion maps,'' {\em Applied and Computational
  Harmonic Analysis}, vol.~21, no.~1, pp.~5--30, 2006.

\bibitem{DBLP:journals/tog/LipmanRF10}
Y.~Lipman, R.~M. Rustamov, and T.~A. Funkhouser, ``Biharmonic distance,'' {\em
  {ACM} Trans. Graph.}, vol.~29, no.~3, 2010.

\end{thebibliography}
\end{document}